\documentclass [12pt]{article}
\usepackage{amsmath,amssymb,cite}
\usepackage{float}
\usepackage[section]{placeins}
\usepackage[vcentermath,enableskew]{youngtab}
\usepackage{tabularx}
\usepackage[english]{babel}
\usepackage[dvips]{graphicx}
\usepackage{indentfirst}
\usepackage{epsfig}
\numberwithin{equation}{section}

\begin{document}

\title{A Multitrace Approach to Noncommutative $\Phi_2^4$}

\author{ Badis Ydri \footnote{Email:ydri@stp.dias.ie} \\
 Institute of Physics, BM Annaba University,\\
BP 12, 23000, Annaba, Algeria.\\
}

\maketitle
\abstract{In this article we provide a multitrace analysis of the theory of noncommutative $\Phi^4$ in two dimensions on the fuzzy sphere ${\bf S}^2_{N,\Omega}$, and on the Moyal-Weyl plane  ${\bf R}^{2}_{\theta, \Omega}$, with a non-zero harmonic oscillator term added. The doubletrace matrix model symmetric under $M\longrightarrow -M$ is solved in closed form. An analytical prediction for the disordered-to-non-uniform-ordered phase transition and an estimation of the triple point, from the termination point of the critical boundary, are derived and compared with previous Monte Carlo measurement.}
\section{Introduction}

A scalar phi-four theory on a non-degenerate noncommutative Euclidean spacetime is a matrix model of the form
\begin{eqnarray}
S&=&{\rm Tr}_H\big(a\Phi\Delta {\Phi}+b{\Phi}^2+c{\Phi}^4\big).\label{fundamental}
\end{eqnarray} 
The Laplacian $\Delta$ defines the underlying geometry, i.e. the metric, of the  noncommutative Euclidean spacetime in the sense of  \cite{Connes:1994yd,Frohlich:1993es}. This is a three-parameter model with the following three known phases: 
\begin{itemize}
\item The usual $2$nd order Ising phase transition between disordered $<\Phi>=0$ and uniform ordered $<\Phi>\sim {\bf 1}$ phases. This appears for small values of $c$. This is the only transition observed in commutative phi-four, and thus it can be accessed in a small noncommutativity parameter expansion, using conventional Wilson renormalization group equation \cite{Wilson:1973jj}. See \cite{Ydri:2012nw} for an analysis along this line applied to the $O(N)$ version of the nc phi-four theory.
\item  A matrix transition between disordered $<\Phi>=0$ and non-uniform ordered $<\Phi>\sim \Gamma$ phases with $\Gamma^2={\bf 1}_H$. For a finite dimensional Hilbert space $H$, this transition coincides, for very large values of $c$,  with the $3$rd order transition of the real quartic matrix model, i.e. the model with $a=0$, which occurs at $b=-2\sqrt{Nc}$. In terms of $\tilde{b}=bN^{-3/2}$ and $\tilde{c}=cN^{-2}$ this reads
\begin{eqnarray}
\tilde{b}=-2\sqrt{\tilde{c}}.\label{cl}
\end{eqnarray} 
This is therefore a transition from a one-cut (disc) phase to a two-cut (annulus) phase \cite{Brezin:1977sv,Shimamune:1981qf}.
\item  A transition between uniform ordered  $<\Phi>\sim {\bf 1}_H$ and non-uniform ordered $<\Phi>\sim \Gamma$ phases.  The non-uniform phase, in which translational/rotational invariance is spontaneously broken, is absent in the commutative theory. The non-uniform phase is essentially  the stripe phase observed originally on  Moyal-Weyl spaces in \cite{Gubser:2000cd,Ambjorn:2002nj}. 
\end{itemize}

Let us discuss a little further the phase structure of the pure potential model $V={\rm Tr}_H({b}\Phi^2+{c}\Phi^4)$, in the case when the Hilbert space $H$ is $N-$dimensional, in some more detail. The ground state configurations  are given by the matrices
\begin{eqnarray}
\Phi_0=0.
\end{eqnarray}
\begin{eqnarray}
\Phi_{\gamma}=\sqrt{-\frac{b}{2c}}U\gamma U^+~,~{\gamma}^2={\bf 1}_N~,~UU^+=U^+U={\bf 1}_N.
\end{eqnarray}
We compute $V[\Phi_0]=0$ and $V[\Phi_{\gamma}]=-b^2/4c$. The first configuration corresponds to the disordered phase characterized by $<\Phi>=0$. The second solution makes sense only for $b<0$, and it corresponds to the ordered phase characterized by $<\Phi>\ne 0$. As mentioned above, there is a non-perturbative transition between the two phases which occurs quantum mechanically, not at $b=0$, but at $b=b_*=-2\sqrt{Nc}$, which is known as the one-cut to two-cut transition. The idempotent $\gamma$ can always be chosen such that $\gamma=\gamma_k={\rm diag}({\bf 1}_{k},-{\bf 1}_{N-k})$. The orbit of $\gamma_k$ is the Grassmannian manifold $U(N)/(U(k)\times U(N-k))$ which is $d_k-$dimensional where  $d_k=2kN-2k^2$. It is not difficult to show that this dimension is maximum at $k=N/2$, assuming that $N$ is even, and hence from entropy argument, the most important two-cut solution is the so-called stripe configuration given by $\gamma={\rm diag}({\bf 1}_{{N}/{2}},-{\bf 1}_{{N}/{2}})$. In this real quartic matrix model, we have therefore three possible phases characterized by the following order parameters:
\begin{eqnarray}
&&<\Phi>=0~~{\rm disordered}~{\rm phase}.
\end{eqnarray}
\begin{eqnarray}
&&<\Phi>=\pm\sqrt{-\frac{b}{2c}}{\bf 1}_N~~{\rm Ising}~({\rm uniform})~{\rm phase}.
\end{eqnarray}
\begin{eqnarray} 
&&<\Phi>=\pm\sqrt{-\frac{b}{2c}}\gamma~~{\rm matrix}~({\rm nonuniform}~{\rm or}~{\rm stripe})~{\rm phase}.
\end{eqnarray}
However, as one can explicitly check by calculating the free energies of the respective phases, the uniform ordered phase is not  stable in the real quartic matrix model  $V={\rm Tr}_H({b}\Phi^2+{c}\Phi^4)$.

\begin{figure}[htbp]
\begin{center}
\includegraphics[width=5.0cm,angle=-90]{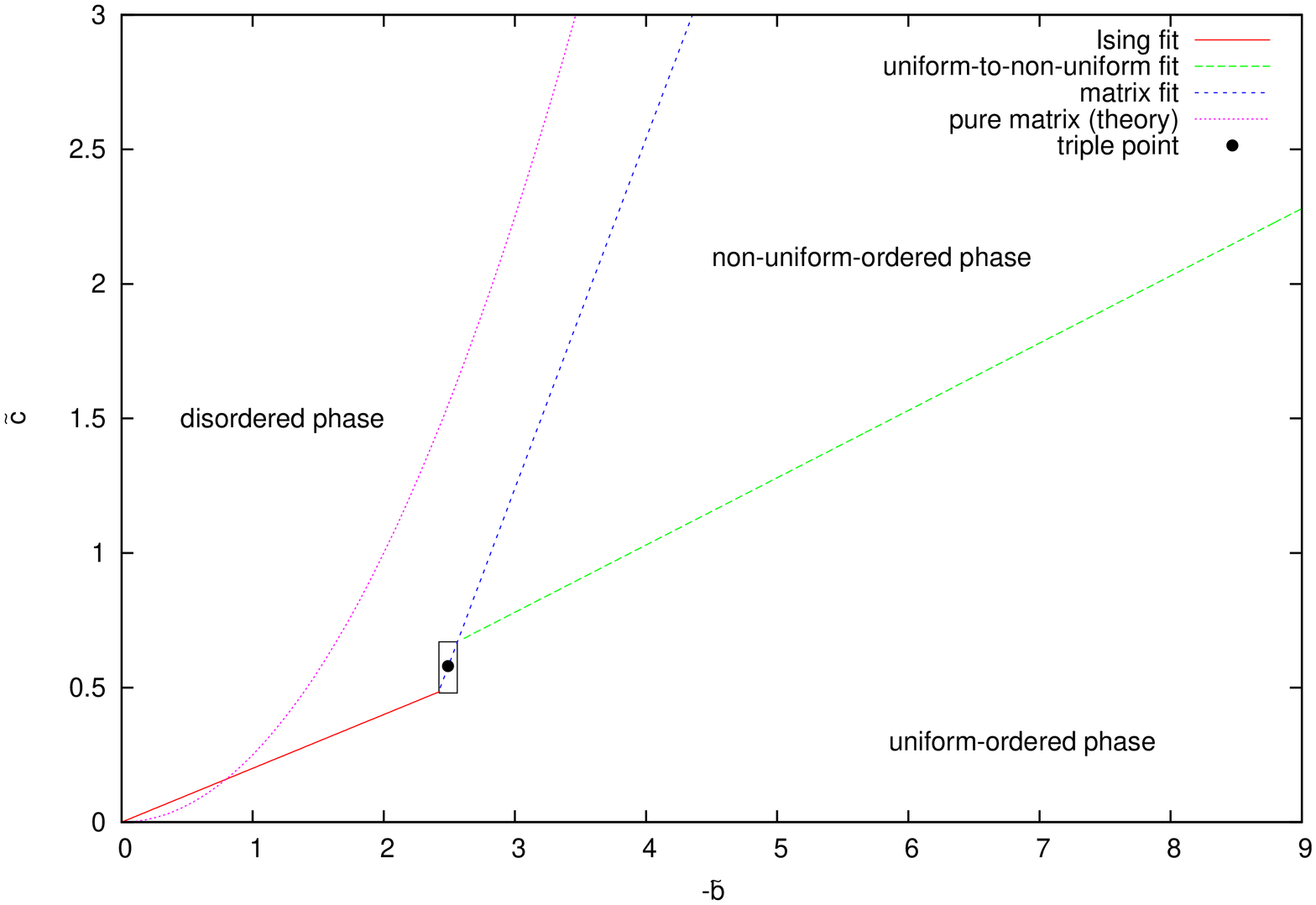}
\includegraphics[width=5.0cm,angle=-90]{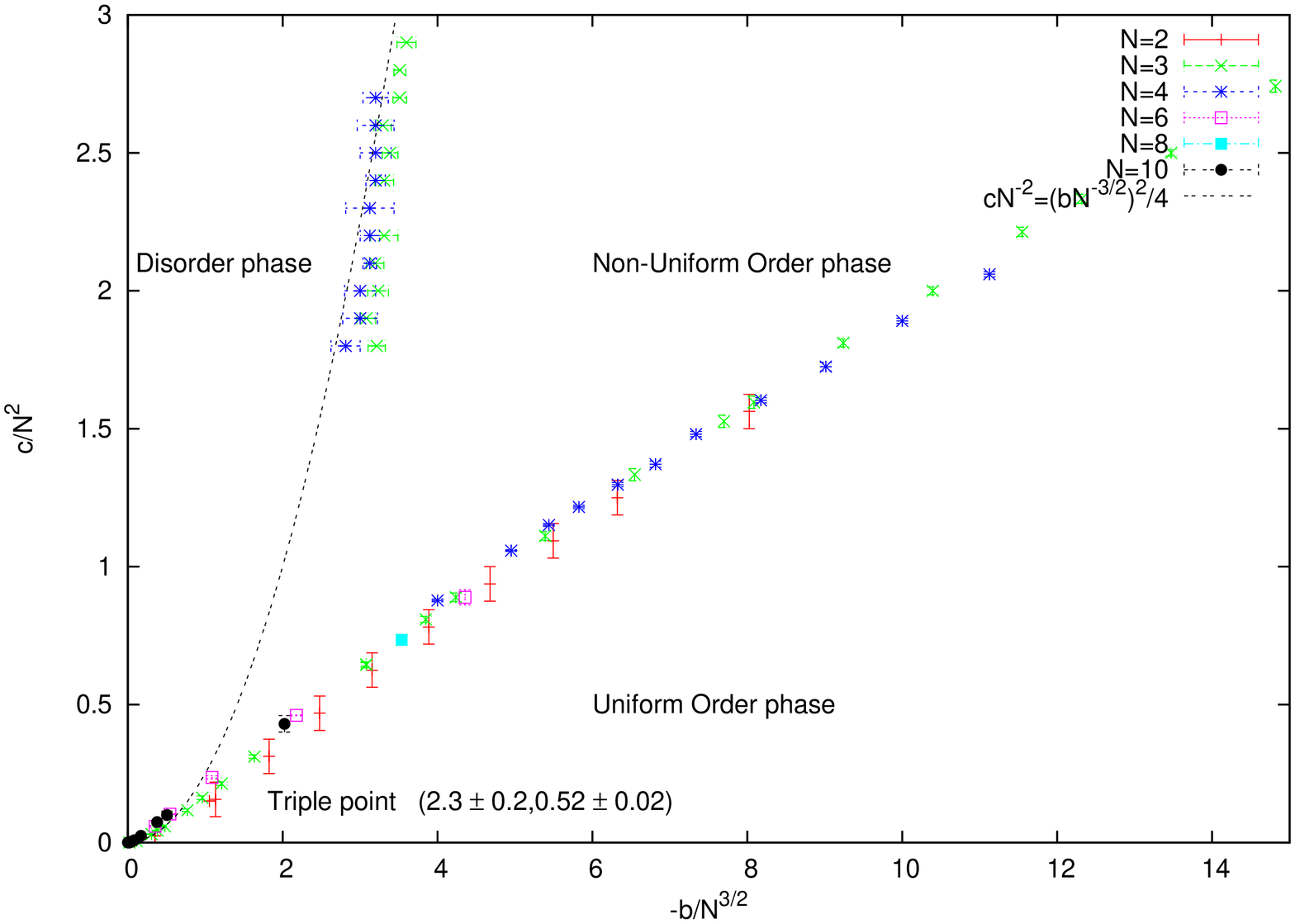}
\caption{The phase diagram of phi-four theory on the fuzzy sphere. In the first figure the fits are reproduced from actual Monte Carlo data \cite{Ydri:2014rea}. Second figure reproduced from \cite{GarciaFlores:2009hf} with the gracious permission of  D.~O'Connor.}\label{phase_diagram}
\end{center}
\end{figure}

The above picture is expected to hold for noncommutative/fuzzy phi-four theory in any dimension, and the three phases are all stable and are expected to meet at a triple point. This structure was confirmed in two dimensions by means of Monte Carlo simulations on the fuzzy sphere in  \cite{GarciaFlores:2009hf,GarciaFlores:2005xc}. The phase diagram is shown on figures (\ref{phase_diagram}). Both figures were generated using the Metropolis algorithm on the fuzzy sphere. In the first figure coupling of the scalar field $\Phi$ to a U(1) gauge field on the fuzzy sphere is included, and as a consequence, we can employ the U(N) gauge symmetry to reduce the scalar sector to only its eigenvalues. 


The problem of the phase structure of fuzzy scalar phi-four was also studied in \cite{Martin:2004un,Panero:2006bx,Das:2007gm,Medina:2007nv}. The analytic derivation of the phase diagram of noncommutative phi-four on the fuzzy sphere was attempted in \cite{O'Connor:2007ea,Saemann:2010bw,Polychronakos:2013nca,Tekel:2014bta}. The related problem of Monte Carlo simulation of noncommutative phi-four on the fuzzy torus, and the fuzzy disc was considered in \cite{Ambjorn:2002nj}, \cite{Bietenholz:2004xs}, and \cite{Lizzi:2012xy} respectively. For a recent study see \cite{Mejia-Diaz:2014lza}.

In this paper, we are interested in studying, by means of the multi-trace approach initiated in \cite{O'Connor:2007ea}, the theory of noncommutative $\Phi^4$ in two dimensions on the fuzzy sphere ${\bf S}^2_{N,\Omega}$ and the Moyal-Weyl plane  ${\bf R}^{2}_{\theta, \Omega}$, with a non-zero harmonic oscillator term.  The construction of the harmonic oscillator term on the Moyal-Weyl plane can be found in \cite{Langmann:2002cc}, whereas the analogue construction on the fuzzy sphere is done in \cite{Ydri:2014rea}. The  multi-trace expansion is the analogue of the Hopping parameter expansion on the lattice in the sense that we perform a small kinetic term expansion, i.e. expanding in the parameter $a$ of (\ref{fundamental}), as opposed to the small potential expansion of the usual perturbation theory \cite{Montvay:1994cy,Smit:2002ug}. This technique is expected to capture the matrix transition between disordered $<\Phi>=0$ and non-uniform ordered $<\Phi>\sim \gamma$ phases with arbitrarily increasing accuracy by including more and more terms in the expansion in $a$. From this we can  then infer and/or estimate the position of the triple point. Capturing the Ising transition requires, in our opinion, the whole expansion in $a$, or at least a very large number of terms in the expansion. This is because, it is not obvious how does a small number of terms in the expansion in $a$ approximates the geometry encoded in the kinetic term, and as a consequence, the Ising phase $<\Phi>\sim {\bf 1}$ will more likely be seen as metastable within this scheme. There is, of course, the expectation that the uniform ordered phase will become stable at some order of this approximation.  

This article is organized as follows:
\begin{itemize}
\item{}Section $2$: The Model and The Method.
\item{}Section $3$: The Real Multitrace Quartic Matrix Model.
\item{}Section $4$: Matrix Model Solutions.
\item{}Section $5$: Monte Carlo Results
\begin{itemize}
\item{}Summary of Models and Algorithm.
\item{}Monte Carlo Tests of Multitrace Approximations.
\item{}Phase Diagrams and Other Physics.
\end{itemize}
\item{}Section $6$: The Nonperturbative Effective Potential Approach.
\item{}Section $7$: Conclusion.
\end{itemize}
We also include three appendices for the benefit of interested readers.

\section{The Model and The Method}
The model studied in this paper, on the fuzzy sphere   ${\bf S}^2_{N,\Omega}$ and on the {regularized} Moyal-Weyl plane  ${\bf R}^{2}_{\theta, \Omega}$, can be rewritten coherently as the following matrix model

\begin{eqnarray}
S[M]&=&r^2K[M]+Tr\big[b M^2+c M^4\big].\label{fundamentale}
\end{eqnarray}
\begin{eqnarray}
K[M]&=&Tr\bigg[\sqrt{\omega}\Gamma^+M\Gamma M-\frac{\epsilon}{N+1}\Gamma_3M\Gamma_3M+EM^2\bigg].
\end{eqnarray}
The first term is precisely the kinetic term. The parameter $\epsilon$ takes one of two possible values corresponding to the topology/metric of the underlying geometry, viz
\begin{eqnarray}
&&\epsilon=1~,~{\rm sphere}\nonumber\\
&&\epsilon=0~,~{\rm plane}.
\end{eqnarray}
The parameters $b$, $c$, $r^2$ and $\sqrt{\omega}$ are related to the mass parameter $m^2$, the quartic coupling constant $\lambda$, the noncommutativity parameter $\theta$ and the harmonic oscillator parameter $\Omega$, of the original model, by the equations
\begin{eqnarray}
b=\frac{1}{2}m^2~,~c=\frac{\lambda}{4!}\frac{1}{2\pi\theta}~,~r^2=\frac{2 (\Omega^2+1)}{\theta}~,~\sqrt{\omega}=\frac{\Omega^2-1}{\Omega^2+1}.
\end{eqnarray}
The matrices $\Gamma$, $\Gamma_3$ and $E$ are given by

\begin{eqnarray}
(\Gamma_3)_{lm}= l{\delta}_{lm}~,~(\Gamma)_{lm}=\sqrt{(m-1)(1-\epsilon \frac{m}{N+1})}{\delta}_{lm-1}~,~(E)_{lm}=(l-\frac{1}{2}){\delta}_{lm}.
\end{eqnarray}
Let us discuss the connection between the actions (\ref{fundamental}) and (\ref{fundamentale}). We note first that the original action (\ref{fundamental}) on the fuzzy sphere, with a non zero harmonic oscillator term, is defined by the Laplacian  \cite{Ydri:2014rea} 
\begin{eqnarray}
\Delta=[L_a,[L_a,...]]+\Omega^2[L_3,[L_3,...]]+\Omega^2\{L_i,\{L_i,...\}\}.
\end{eqnarray}
Explicitly we have
\begin{eqnarray}
S=\frac{4\pi R^2}{N+1} Tr\bigg(\frac{1}{2R^2}{\Phi}\Delta{\Phi}+\frac{1}{2}m^2{\Phi}^2+\frac{\lambda}{4!}{\Phi}^4\bigg).
\end{eqnarray}
Equivalently this action with the substitution ${\Phi}={\cal M}/\sqrt{2\pi\theta}$, where ${\cal M}=\sum_{i,j=1}^NM_{ij}|i><j|$, reads 
\begin{eqnarray}
S=Tr\bigg(a{\cal M}\Delta_{N,\Omega}{\cal M}+b{\cal M}^2+c{\cal M}^4\bigg)\label{fundamentale2}.
\end{eqnarray}
This is  is identical  to  (\ref{fundamentale}). The relationship between the parameters $a=1/(2R^2)$ \footnote{The noncommutativity parameter on the fuzzy sphere is related to the radius of the sphere by $\theta=2R^2/\sqrt{N^2-1}$.} and $r^2$ is given by  $r^2=2a(\Omega^2+1)N$.

We start from the path integral  

\begin{eqnarray}
Z=\int d M ~\exp\big(-S[M]\big).
\end{eqnarray}
First, we will diagonalize the scalar matrix as $M=U\Lambda U^{-1}$. The measure becomes $ d  M= d\Lambda dU \Delta^2(\Lambda)$, where $dU$ is the usual Haar measure over the group SU(N) which is normalized such that $\int dU=1$, whereas the Jacobian $\Delta^2(\Lambda)$ is precisely the so-called Vandermonde determinant defined by $\Delta^2(\Lambda)= \prod_{i>j}(\lambda_i-\lambda_j)^2$.
The path integral becomes the eigenvalues problem
\begin{eqnarray}
Z=\int d \Lambda~\Delta^2(\Lambda) ~\exp\bigg(-Tr\big(b{\Lambda}^2+c{\Lambda}^4\big)\bigg)\int dU~\exp\bigg(-aK[U\Lambda U^{-1}]\bigg).
\end{eqnarray}
The fundamental question we want to answer is: can we integrate the unitary group completely?

The answer, which is the straightforward and obvious one, is to expand the kinetic term in powers of $a=r^2$, perform the integral over $U$, then resume the sum back into an exponential to obtain an effective potential. This is very reminiscent of the hopping parameter expansion on the lattice. This approximation will clearly work if, for whatever reason, the kinetic term is indeed small compared to the potential term which, as it turns out, is true in the matrix phase of noncommutative phi-four theory.

Towards this end, we will take the following steps:
\begin{itemize}
\item{}We expand the scalar field $M$ in the basis formed by the Gell-Mann matrices $t_a$ and the identity matrix $t_0={\bf 1}_N/\sqrt{2N}$ as $M=\sum_{A}M^{A}t_{A}$. The path integral becomes
\begin{eqnarray}
Z=\int  d\Lambda  \Delta^2(\Lambda)\exp\bigg(-Tr\big(b \Lambda^2+c \Lambda^4\big)\bigg)\int dU ~\exp\big(-a K_{AB}(Tr U\Lambda U^{-1}t_{A})(Tr U\Lambda U^{-1}t_{B})\big).\nonumber\\
\end{eqnarray}
The  kinetic matrix $K$ is given explicitly by 
\begin{eqnarray}
K_{AB}&=&2\sqrt{\omega}Tr\Gamma^+t_A\Gamma t_B+2\sqrt{\omega}Tr\Gamma^+t_B\Gamma t_A-\frac{4\epsilon}{N+1}Tr\Gamma_3t_A\Gamma_3t_B+2TrE\{t_A,t_B\}.\label{KAB}\nonumber\\
\end{eqnarray} 
\item{}We expand in powers of $a$. In this paper we only go  upto the second order in $a$. This can be extended to any order in an obvious way as we will see.
\item{}We use  $(Tr A)(TrB)=Tr_{N^2}(A\otimes B)$ and $(A\otimes C)(B\otimes D)=AB\otimes CD$. 
\item{}We decompose the  $N^2-$dimensional and the $N^4-$dimensional  Hilbert spaces, under the SU(N) action, into the direct sums of  subspaces corresponding to the irreducible representations $\rho$ contained in $N\otimes N$ and $N\otimes N\otimes N\otimes N$ respectively.  The tensor products of interest are
\begin{eqnarray}
\young(A)\otimes \young (B)=\young(AB)\oplus \young(A,B).
\end{eqnarray}
\begin{eqnarray}
\young(A)\otimes \young(B)\otimes \young(C)\otimes \young(D)&=&\young(ABCD)\oplus\young(A,B,C,D)\oplus \young(ABC,D)\oplus \young(ABD,C)\oplus \young(ACD,B)\nonumber\\
&\oplus & \young(AD,B,C)\oplus \young(AC,B,D)\oplus \young(AB,C,D)\oplus \young(AB,CD)\oplus \young(AC,BD).
\end{eqnarray}
\item{}We use the orthogonality relation 
\begin{eqnarray}
\int dU \rho(U)_{ij}\rho(U^{-1})_{kl}=\frac{1}{{\rm dim}(\rho)}\delta_{il}\delta_{jk}.
\end{eqnarray}
We obtain [see \cite{O'Connor:2007ea,Saemann:2010bw} and \cite{Ydri_Lectures} for more detail]

\begin{eqnarray}
\int dU~\exp\bigg(-aK[U\Lambda U^{-1}] \bigg)
&=&1-aK_{AB}\sum_{\rho}\frac{1}{{\rm dim}(\rho)}Tr_{\rho}\Lambda\otimes\Lambda. Tr_{\rho}t_A\otimes t_B\nonumber\\
&+&\frac{1}{2!}a^2K_{AB}K_{CD}\sum_{\rho}\frac{1}{{\rm dim}(\rho)}Tr_{\rho}\Lambda\otimes\Lambda. Tr_{\rho}t_A\otimes...\otimes t_D+...\nonumber\\
\end{eqnarray}
Thus, the calculation of the first and second order corrections reduce to the calculation of the traces $Tr_{\rho}t_A\otimes t_B$ and $Tr_{\rho}t_A\otimes t_B\otimes t_C\otimes t_D$ respectively. It is then obvious, that generalization to higher order corrections will involve the traces  $Tr_{\rho}t_{A_1}\otimes ...\otimes t_{A_n}$ and  $Tr_{\rho}\Lambda\otimes ...\otimes \Lambda$. Explicitly the $n$th order correction should read
\begin{eqnarray}
n{\rm th}~{\rm order}
&=&\frac{1}{n!}K_{A_1A_2}...K_{A_{2n-1}A_{2n}}\sum_{\rho}\frac{1}{{\rm dim}(\rho)}Tr_{\rho}\Lambda\otimes... \otimes \Lambda.Tr_{\rho}t_{A_1}\otimes t_{A_2}\otimes...\otimes t_{A_{2n-1}}\otimes t_{A_{2n}}.\nonumber\\
\end{eqnarray}
\item{}By substituting the dimensions of the various irreducible representations and the relevant SU(N) characters we arrive at the formula (with $t_i=Tr\Lambda^i$)

\begin{eqnarray}
\int dU~\exp\bigg(-aK[U\Lambda U^{-1}]\bigg)
&=&1-\frac{a}{2}\bigg[(s_{1,2}+s_{2,1})t_1^2+(s_{1,2}-s_{2,1})t_2\bigg]\nonumber\\
&+&\frac{a^2}{2}\bigg[\frac{1}{4}(s_{1,4}-s_{4,1}-s_{2,3}+s_{3,2})t_4+\frac{1}{3}(s_{1,4}+s_{4,1}-s_{2,2})t_1t_3\nonumber\\
&+&\frac{1}{8}(s_{1,4}+s_{4,1}-s_{2,3}-s_{3,2}+2s_{2,2})t_2^2+\frac{1}{4}(s_{1,4}-s_{4,1}+s_{2,3}-s_{3,2})t_2t_1^2\nonumber\\
&+&\frac{1}{24}(s_{1,4}+s_{4,1}+3s_{2,3}+3s_{3,2}+2s_{2,2})t_1^4\bigg]+....
\end{eqnarray}
\item{}There remains the explicit calculation of the coefficients $s$, taking the large $N$ limit, and finally re-exponetiating the series back to obtain the effective potential. This is a considerably long calculation which is done originally on the fuzzy sphere in  \cite{O'Connor:2007ea}, and extended to the current case, which includes a harmonic oscillator term, in  \cite{Ydri_Lectures}. We will skip here the lengthy detail.

The definition of the coefficients $s$ in terms of the kinetic matrix $K$ and characters and dimensions of various SU(N)/U(N) representations,  the explicit calculation of these coefficients as well as extraction of the large $N$ behavior are sketched in the appendix.
\end{itemize}

\section{The Real Multitrace Quartic Matrix Model}

The end result of the above steps  is the effective potential \cite{Ydri_Lectures}


\begin{eqnarray}
 \Delta V&=&\frac{r^2}{4}\bigg(v_{2,1}T_2+\frac{2N}{3}w_1t_2\bigg)+\frac{r^4}{24}\bigg(v_{4,1}T_4-\frac{4}{N^2}v_{2,2} T_2^2+4w_2(t_1 t_3-t_2^2)+\frac{4}{N}w_3 t_2T_2\bigg)+O(r^6).\label{endresult}\nonumber\\
\end{eqnarray}
The complete effective action in terms of the eigenvalues is the sum of the classical potential, the Vadermonde determinant and  the above effective potential.  The operators $T_2$ and $T_4$ are defined below. The coefficients $v$ and $w$ are given by \cite{Ydri_Lectures}

 \begin{eqnarray}
 v_{2,1}=2-\epsilon-\frac{2}{3}(\sqrt{\omega}+1)(3-2\epsilon).
\end{eqnarray}
\begin{eqnarray}
 v_{4,1}=-(1-\epsilon).
\end{eqnarray}
 \begin{eqnarray}
 v_{2,2}
&=&w_3+(\sqrt{\omega}+1)(1-\epsilon)+\frac{1}{12}({\omega}-1)(9-8\epsilon)-\frac{1}{8}(2-3\epsilon).
\end{eqnarray}
\begin{eqnarray}
w_1&=&(\sqrt{\omega}+1)(3-2\epsilon).
\end{eqnarray}
\begin{eqnarray}
w_2&=&-(\sqrt{\omega}+1)(1-\epsilon).
\end{eqnarray}
\begin{eqnarray}
w_3
&=&(\sqrt{\omega}+1)(1-\epsilon)-\frac{1}{15}(\sqrt{\omega}+1)^2(15-14\epsilon).
\end{eqnarray}

Three important remarks are now in order:
\begin{itemize}
\item{}{\bf Zero Mode:}
We know that, in the limit $\Omega^2\longrightarrow 0$ ($\sqrt{\omega}\longrightarrow -1$), the trace part of the scalar field drops from the kinetic action, and as a conseqeunce, the effective potential can be rewritten solely in terms of the differences $\lambda_i-\lambda_j$ of the eigenvalues. Furthermore, in this limit, the effective potential must also be invariant under any permutation of the eigenvalues, as well as under the parity $\lambda_i\longrightarrow -\lambda_i$, and hence it can only depend on the following functions \cite{O'Connor:2007ea}

\begin{eqnarray}
T_4&=&Nt_4-4t_1t_3+3t_2^2=\frac{1}{2}\sum_{i\neq j}(\lambda_i-\lambda_j)^4.
\end{eqnarray}
\begin{eqnarray}
T_2^2&=&\bigg[\frac{1}{2}\sum_{i\neq j}(\lambda_i-\lambda_j)^2\bigg]^2=t_1^4-2Nt_1^2t_2+N^2t_2^2.
\end{eqnarray}
\begin{eqnarray}
T_n^m&=&\bigg[\frac{1}{2}\sum_{i\neq j}(\lambda_i-\lambda_j)^n\bigg]^m.
\end{eqnarray}
It is clear that only the functions $T_2$ and $T_4$ can appear at the second order in $a=r^2$. We also observe that the quadratic part of the resulting effective potential can be expressed, modulo a term which vanishes as  $\sqrt{\omega}+1$ in the limit $\sqrt{\omega}\longrightarrow -1$, in terms of the function 
\begin{eqnarray}
T_2&=&Nt_2-t_1^2=\frac{1}{2}\sum_{i\neq j}(\lambda_i-\lambda_j)^2.
\end{eqnarray}
In general, it is expected that for generic values of $\sqrt{\omega}$, away from the zero harmonic oscillator case $\sqrt{\omega}=-1$, the effective potential will contain terms proportional to $\sqrt{\omega}+1$ which can not be expressed solely in terms of the functions $T_2$, $T_4$, etc. This is obvious from the result (\ref{endresult}).
\item{}{\bf The Case $\epsilon=1$, $\Omega^2=0$:} In this case $v_{2,1}=1$, $v_{4,1}=0$, $v_{2,2}=1/8$ while all the $w$ coefficients vanish. We get then the effective potential

\begin{eqnarray}
 \Delta V&=&\frac{r^2}{4}T_2-\frac{r^4}{48N^2} T_2^2+O(r^6).\label{our}
\end{eqnarray}
This result is very different from the one obtained in  \cite{O'Connor:2007ea}\footnote{Compare with equation  $(4.4)$  of \cite{O'Connor:2007ea}.} in which the coefficient $v_{4,1}$ was found to be non zero, more precisely $v_{4,1}=3/2$, while the correction associated with the coefficient $v_{2,2}$ was suppressed, i.e. they set $v_{2,2}=0$, and hence the effective potential, in their case, is given by
\begin{eqnarray}
 \Delta V&=&\frac{r^2}{4}T_2+\frac{r^4}{16} T_4+O(r^6).\label{theirs}
\end{eqnarray}
A detailed discussion of this point can be found in \cite{Ydri_Lectures}, while a concise description of the discrepancy is included in the appendix. However, we should note that although these two results are quantitatively  different the resulting physics is  qualitatively the same.

\item{}{\bf Scaling:} 
From the Monte Carlo results of \cite{GarciaFlores:2009hf,GarciaFlores:2005xc} on the fuzzy sphere, we know that the scaling behavior of the parameters $a$, $b$ and $c$ appearing in the action (\ref{fundamentale2}) is given by
\begin{eqnarray}
\bar{a}=\frac{a}{N^{\delta_a}}~,~\bar{b}=\frac{b N^{2\delta_{\lambda}}}{aN^{3/2}}~,~\bar{c}=\frac{cN^{4\delta_{\lambda}}}{a^2N^2}.\label{mc}
\end{eqnarray}
In the above equation we have also included a possible scaling of the field/matrix $M$, which is not included in \cite{GarciaFlores:2009hf,GarciaFlores:2005xc}, given by $\delta_{\lambda}$. The scaling of the parameter $a$ encodes the scaling of the radius $R^2$ or equivalently the noncommutativity parameter $\theta$. There is of course an extra parameter in  the current case given by  $d=a\Omega^2$, or equivalently $\sqrt{\omega}=(\Omega^2-1)/(\Omega^2+1)$, which comes with another scaling $\delta_{d}$ not discussed altogether in Monte Carlo simulations.

We will assume, in most of this paper, that the four parameters $b$, $c$, $r^2$ and $\sqrt{\omega}$ of the matrix model (\ref{fundamentale}) scale as
 \begin{eqnarray}
\tilde{b}=\frac{b}{N^{\delta_{b}}}~,~\tilde{c}=\frac{c}{N^{\delta_{c}}}~,~\tilde{r}^2=\frac{r^2}{N^{\delta_{r}}}~,~\sqrt{\tilde{\omega}}=\frac{\sqrt{\omega}}{N^{\delta_{\omega}}}.
\end{eqnarray}
Obviously $\delta_r=\delta_a+1$. Further, we will assume a scaling $\delta_{\lambda}$ of the  eigenvalues $\lambda$, viz
 \begin{eqnarray}
\tilde{\lambda}=\frac{\lambda}{N^{\delta_{\lambda}}}.
\end{eqnarray}
Hence, in order for the effective action to come out of order $N^2$, we must  have the following values
 \begin{eqnarray}
\delta_{b}=1-2\delta_{\lambda}~,~\delta_{c}=1-4\delta_{\lambda}~,~\delta_{r}=-2\delta_{\lambda}~,~\delta_{\omega}=0.
\end{eqnarray}
 By substituting in (\ref{mc}) we obtain the collapsed exponents
\begin{eqnarray}
\delta_{\lambda}=-\frac{1}{4}~,~\delta_a=-\frac{1}{2}~,~\delta_{b}=\frac{3}{2}~,~\delta_{c}=2~,~\delta_{d}=-\frac{1}{2}~,~\delta_{r}=\frac{1}{2}.
\end{eqnarray}
In simulations, it is found that the scaling behavior of the mass parameter $b$ and the quartic coupling $c$ is precisely given by $3/2$ and $2$ respectively. We will assume, for simplicity, the same scaling on the Moyal-Weyl plane. 

\end{itemize}

\section{Matrix Model Solutions}
The saddle point equation  corresponding to the sum $V_{r^2,\Omega}$ of the classical potential and the effective potential (\ref{endresult}), which also includes the appropriate scaling \footnote{The eigenvalues here are also scaled only we suppress the tilde for ease of notation.}, takes the form
\begin{eqnarray}
\frac{1}{N} S_{\rm eff}^{'} &=&V_{r^2,\Omega}^{'}-\frac{2}{N}\sum_i\frac{1}{{\lambda}-{\lambda}_i}\nonumber\\
&=&0.\label{sp1e}
\end{eqnarray}
Next, we will assume
a symmetric support of the eigenvalues distributions, and as a consequence, all odd moments vanish identically  \cite{O'Connor:2007ea}. This is motivated by the fact that the expansion of the effective action employed in the current paper, i.e. the multitrace technique,  is expected to probe, very well, the transition between the disordered phase and the non-uniform ordered phase. 

We will, therefore,  assume here that across the transition line between disordered phase and non-uniform ordered phase, the matrix $M$ remains massless, and the eigenvalues distribution $\rho({\lambda})$ is always symmetric, and hence all odd moments ${m}_q$ vanish identically, viz
   \begin{eqnarray} 
{m}_q=\int_a^b d{\lambda} \rho({\lambda}){\lambda}^q=0~,~q={\rm odd}.
\end{eqnarray}
The derivative of the generalized potential $V_{r^2,\Omega}^{'}$ is therefore given by 
\begin{eqnarray}
V_{r^2,\Omega}^{'}({\lambda})&=&2\tilde{b}{\lambda}+4\tilde{c}{\lambda}^3+\tilde{r}^2(\frac{v_{2,1}}{2}+\frac{w_1}{3}){\lambda}\nonumber\\
&+&\tilde{r}^4\bigg[\frac{1}{6}v_{4,1}\big({\lambda}^3+3{m}_2{\lambda}\big)-\frac{2}{3}w_2{m}_2{\lambda}-\frac{2}{3}v_{2,2} {m}_2{\lambda}+\frac{1}{3}w_{3}{m}_2{\lambda}\bigg]+...\nonumber\\
\end{eqnarray}
The corresponding matrix model potential and effective action are given respectively by the following
\begin{eqnarray}
V_{r^2,\Omega}
&=&N\int d\lambda \rho(\lambda)\bigg[\bigg(\tilde{b}+\frac{\tilde{r}^2}{2}\big(\frac{v_{2,1}}{2}+\frac{w_1}{3}\big)\bigg){\lambda}^2+\big(\tilde{c}+\frac{\tilde{r}^4}{24}v_{4,1}\big){\lambda}^4\bigg]-\frac{\tilde{r}^4N}{6}\eta\bigg[\int d\lambda \rho(\lambda){\lambda}^2\bigg]^2.\nonumber\\
\end{eqnarray}
\begin{eqnarray}
S_{\rm eff}=NV_{r^2,\Omega}-\frac{N^2}{2}\int d\lambda d\lambda^{'}\rho(\lambda)\rho(\lambda^{'})\ln (\lambda-\lambda^{'})^2.
\end{eqnarray}
The coefficient $\eta$ is defined by
\begin{eqnarray}
\eta&=&v_{2,2}-\frac{3}{4}v_{4,1}+w_2-\frac{1}{2}w_3\nonumber\\
&=&\frac{1}{8}(4-3\epsilon)-\frac{1}{6}(\sqrt{\omega}+1)(6-5\epsilon)+\frac{1}{20}(\sqrt{\omega}+1)^2(5-4\epsilon).
\end{eqnarray}
These can be derived from the matrix model given by
\begin{eqnarray}
V_{r^2,\Omega}&=&\mu_0 Tr M^2+g_0 Tr M^4-\frac{\tilde{r}^4}{6N}\eta\bigg[ Tr M^2\bigg]^2.\label{mtmm}
\end{eqnarray}
The parameters $\mu_0$ and $g_0$ are defined by
\begin{eqnarray}
&&\mu_0=\tilde{b}+\frac{\tilde{r}^2}{2}\big(\frac{v_{2,1}}{2}+\frac{w_1}{3}\big)=\tilde{b}+\frac{\tilde{r}^2}{4}(2-\epsilon)~,~\nonumber\\
&&g_0=\tilde{c}+\frac{\tilde{r}^4}{24}v_{4,1}=\tilde{c}-\frac{\tilde{r}^4}{24}(1-\epsilon).
\end{eqnarray}
This matrix model was studied originally in \cite{Das:1989fq} within the context of $c>1$ string theories. The dependence of this result on the harmonic oscillator potential is fully encoded in the parameter $\eta$ which is the strength of the double trace term since $\mu_0$ and $g_0$ are independent of $\sqrt{\omega}$. For later purposes we rewrite the derivative of the generalized potential $V_{r^2,\Omega}^{'}$ in the suggestive form

\begin{eqnarray}
V_{r^2,\Omega}^{'}({\lambda})&=&2\mu{\lambda}+4g{\lambda}^3.\label{sppp}
\end{eqnarray}
\begin{eqnarray}
&&\mu=\mu_0-\frac{\tilde{r}^4}{3}\eta{m}_2~,~g=g_0.
\end{eqnarray}
The above  saddle point equation (\ref{sp1e}) can be solved using the approach outlined in \cite{eynard} for real single trace quartic matrix models. We only need to account here for the fact that the mass parameter $\mu$ depends on the eigenvalues through the second moment $m_2$. In other words, besides the normalization condition which the eigenvalues distribution must satisfy, we must also satisfy the requirement that the computed second moment $m_2$, using this eigenvalues density, will depend on the mass parameter $\mu$ which itslef is a function of the second moment $m_2$.

The phase structure of the  real quartic matrix model is described concisely in \cite{Shimamune:1981qf}. The two stable phases of the theory are the one-cut (disk)  and the two-cut (annulus) phases which are separated by the critical line (\ref{cl})\footnote{At $\tilde{b}=-2\sqrt{\tilde{c}}$, the eigenvalues density approaches the same behavior from both sides of the transition. This is the sense in which this phase transition is termed critical although it is actually $3$rd order as seen from the behavior of the specific heat.}. There exists also an asymmetric (uniform) one-cut solution which corresponds to a metastable phase. Here, for our real multitrace quartic matrix model (\ref{mtmm}), the phase diagram will consist of the same stable phases, separated by a deformation of the critical line   (\ref{cl}), as well as an analogous metastable  asymmetric one-cut phase.  

In the remainder, we discuss further the two stable phases of the real multitrace quartic matrix model (\ref{mtmm}), the critical boundary between them, as well as a lower estimation of the triple point. More detail can be found in \cite{Ydri_Lectures}.

\paragraph{The Disordered Phase:} The one-cut (disk) solution is given by the equation
\begin{eqnarray}
\rho(\lambda)=\frac{1}{\pi}(2g_0{\lambda}^2+\frac{2}{\delta^2}-\frac{g_0\delta^2}{2})\sqrt{\delta^2-{\lambda}^2}.
\end{eqnarray}
The radius $\delta^2=x$ is the solution of a depressed quartic equation given by

\begin{eqnarray}
\tilde{r}^4\eta g_0x^4-72(g_0-\frac{\tilde{r}^4}{18}\eta)x^2-48\mu_0 x+96=0.\label{quex}
\end{eqnarray}
This eigenvalues distribution is always positive definite for
  \begin{eqnarray}
x^2\leq x_*^2=\frac{4}{g_0}.
\end{eqnarray}
Obviously, $x_*$ must also be a solution of the quartic equation (\ref{quex}). By substitution, we get the solution
\begin{eqnarray}
\mu_{0*} =-2\sqrt{g_0}+\frac{\eta\tilde{r}^4}{3\sqrt{g_0}}.\label{cl1}
\end{eqnarray}
This critical value $\mu_{0 *}$ is negative for  $g_0\geq {\eta\tilde{r}^4}/{6}$. As expected this line is a deformation of the real quartic matrix model critical line $\mu_{0*} =-2\sqrt{g_0}$. In terms of the original parameters, we have 
\begin{eqnarray}
\tilde{b}_{*} =-\frac{\tilde{r}^2}{4}(2-\epsilon)-2\sqrt{\tilde{c}-\frac{\tilde{r}^4}{24}(1-\epsilon)}+\frac{\eta\tilde{r}^4}{3\sqrt{\tilde{c}-\frac{\tilde{r}^4}{24}(1-\epsilon)}}.
\end{eqnarray}
This result, to our knowledge, is completely new. By assuming that the parameter $\eta$ is positive, the range of this solution is found to be $\mu_0\geq \mu_{0*}$. The second moment ${m}_2$ corresponding to this solution is given by the equation 
\begin{eqnarray}
{m}_2&=&\frac{9g_0}{2\tilde{r}^4\eta x}(x-x_+)(x-x_-).
\end{eqnarray}
\begin{eqnarray}
x_{\pm}=\frac{1}{3g_0}(-\mu_0 \pm \sqrt{{\mu}_0^2+12g_0}).
\end{eqnarray}
This is always positive since $x>x_+>0>x_-$.
\paragraph{The Non-Uniform Ordered Phase:} The two-cut (annulus) solution is given by
\begin{eqnarray}
\rho(\lambda)=\frac{2g_0}{\pi}|{\lambda}|\sqrt{(\lambda^2-\delta_1^2)(\delta_2^2-\lambda^2)}.
\end{eqnarray}
The radii $\delta_1$ and $\delta_2$ are given by
\begin{eqnarray}
\delta_1^2=\frac{3}{6g_0-\tilde{r}^4\eta}(-\mu_0 -2\sqrt{g_0}+\frac{\tilde{r}^4\eta}{3\sqrt{g_0}})~,~\delta_2^2=\frac{3}{6g_0-\tilde{r}^4\eta}(-\mu_0+2\sqrt{g_0}-\frac{\tilde{r}^4\eta}{3\sqrt{g_0}}).
\end{eqnarray}
 We have $\delta_1^2\geq 0$, and by construction then $\delta_2^2\geq \delta_1^2$, iff
\begin{eqnarray}
g_0\geq \frac{\tilde{r}^4\eta}{6}~,~\mu_0\leq \mu_{0*}.
\end{eqnarray}
The critical value $\mu_{0*}$ is still given by (\ref{cl1}), i.e. the range of $\mu$ of this phase meshes exactly with the range of $\mu$ of the previous phase.
\paragraph{The Triple Point:} In the rest of this paper, we will concentrate only on the case of the fuzzy sphere, while we will leave the case of the Moyal-Weyl plane as an exercise. 

In the case of  the fuzzy sphere, i.e. $\epsilon=1$,  we have the following critical line
\begin{eqnarray}
\tilde{b}_{*} =-\frac{\tilde{r}^2}{4}-2\sqrt{\tilde{c}}+\frac{\eta\tilde{r}^4}{3\sqrt{\tilde{c}}}.
\end{eqnarray}
We recall that $r^2=2a(\Omega^2+1)N$ or equivalently $\tilde{r}^2=2\tilde{a}(\Omega^2+1)$. The above critical line in terms of the scaled parameters (\ref{mc}) reads then 
\begin{eqnarray}
\bar{b}_{*} =-\frac{\Omega^2+1}{2}-2\sqrt{\bar{c}}+\frac{4\eta(\Omega^2+1)^2}{3\sqrt{\bar{c}}}.
\end{eqnarray}
This should be compared with (\ref{cl}). The range $g_0\geq {\tilde{r}^4\eta}/{6}$ of this critical line reads now
\begin{eqnarray}
\bar{c}\geq  \frac{2\eta (\Omega^2+1)^2}{3}.\label{rangec}
\end{eqnarray}
The termination point of this line provides a lower estimate of the triple point and it is located at
\begin{eqnarray}
(\bar{b},\bar{c})_T=\bigg(-\frac{\Omega^2+1}{2},  \frac{2\eta (\Omega^2+1)^2}{3}\bigg).
\end{eqnarray}
We have verified numerically the consistency of the above analytic solution extensively. The starting point is the quartic equation (\ref{quex}). We have checked, among other things, that for all $\bar{b}\geq \bar{b}_*$ there exists a positive solution $x$ of (\ref{quex}) which satisfies  $x\leq x_*$ and $x>x_+$, i.e. with positive second moment $m_2$. From the other side, i.e. for  $\bar{b}< \bar{b}_*$, there ceases to exist any solution of (\ref{quex}) with these properties. This behavior extends down until around $\bar{c}_T$. The basics of the algorithm used are explained in the appendix.

\begin{figure}[htbp]
\begin{center}
\includegraphics[width=5.0cm,angle=-90]{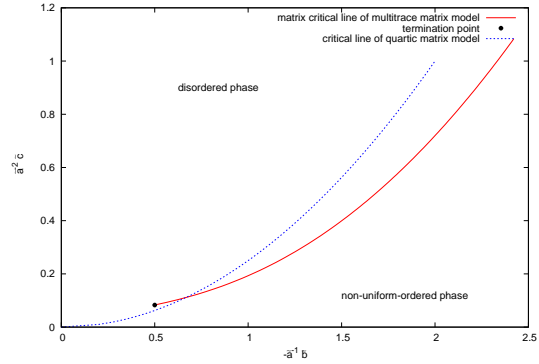}
\caption{The disordered-to-non-uniform-ordered (matrix) transition of phi-four theory on the fuzzy sphere.}\label{clfig}
\end{center}
\end{figure}

Recall that in the case of the fuzzy sphere with a harmonic oscillator term the coefficient $\eta$ is given by
\begin{eqnarray}
\eta
&=&\frac{1}{8}-\frac{1}{6}(\sqrt{\omega}+1)+\frac{1}{20}(\sqrt{\omega}+1)^2.
\end{eqnarray}
For zero harmonic oscillator, i.e. for the ordinary noncommutative phi-four theory on the fuzzy sphere with $\Omega^2=0$ and $\sqrt{\omega}=-1$, we have then the results
\begin{eqnarray}
\bar{b}_{*} =-\frac{1}{2}-2\sqrt{\bar{c}}+\frac{1}{6\sqrt{\bar{c}}}.\label{our2}
\end{eqnarray}
\begin{eqnarray}
\bar{c}\geq  \frac{1}{12}.
\end{eqnarray}
This line is shown on figure (\ref{clfig}). The limit for large $\bar{c}$ is essentially given by (\ref{cl}). As discussed above, the termination point of this line, which is located at

\begin{eqnarray}
(\bar{b}, \bar{c})_T=(-1/2,1/12),\label{term}
\end{eqnarray}
yields a lower estimation of the triple point. This is quite far from the actual value of the triple point found in  \cite{O'Connor:2007ea} to lie at $\sim (-2.3,0.5)$, but it provides an explicit and robust indication that the disordered to non-uniform-ordered transition line does not extend to zero as in the case of real quartic matrix model. 

In any event, the above prediction hinges on the calculated value of the parameter $\eta$ which is expected to increase in value if we include higher order corrections. Furthermore, the inclusion of other multitrace terms, which will arise in higher order calculations, will also affect this result.

It is obvious that the above behavior should hold, essentially unchanged, on the regularized Moyal-Weyl plane.

\section{Monte Carlo Results}
\subsection{Summary of Models and Algorithm}
We start by rewriting the effective action on the fuzzy sphere without a harmonic oscillator term in a form suited for Monte Carlo. The multitrace matrix models of interest are of the form 
\begin{eqnarray}
V&=&V_0+\Delta V=V_0+V_2+V_4.
\end{eqnarray}
The quartic matrix model $V_0$ and the quadratic and quartic corrections $V_2$ and $V_4$ are given explicitly by 
\begin{eqnarray}
V_0&=&{b}Tr M^2+{c}Tr M^4.
\end{eqnarray}
\begin{eqnarray}
V_2=F^{'} Tr M^2+B^{'} (Tr M)^2.
\end{eqnarray}
\begin{eqnarray}
V_4&=&E^{'} Tr M^4+C^{'} Tr M Tr M^3+D^{'}(Tr M)^4+A^{'}Tr M^2 (Tr M)^2+D\big(Tr M^2\big)^2.\nonumber\\
\end{eqnarray}
The primed parameters of the model are
\begin{eqnarray}
&&F^{'}=\frac{aN^2v_{2,1}}{2}~,~B^{'}=-\frac{aN}{2}v_{2,1}\nonumber\\
&&E^{'}=\frac{a^2N^3v_{4,1}}{6}~,~C^{'}=-\frac{2a^2N^2}{3}v_{4,1}~,~D^{'}=-\frac{2a^2}{3}v_{2,2}~,~A^{'}=\frac{4a^2N}{3}v_{2,2}. 
\end{eqnarray}
The remaining parameter $D$ is given by
\begin{eqnarray}
D=-\frac{2a^2N^2}{3}\eta~,~\eta=v_{2,2}-\frac{3}{4}v_{4,1}.
\end{eqnarray}
The parameters $F^{'}$ and $E^{'}$ can be reabsorbed into $b$ and $c$ as

\begin{eqnarray}
B={b}+\frac{aN^2v_{2,1}}{2}~,~C={c}+\frac{a^2N^3v_{4,1}}{6}.
\end{eqnarray}
The effective action we want to study becomes
\begin{eqnarray}
V
&=&Tr\big(B M^2+C M^4\big)+D\big(Tr M^2\big)^2\nonumber\\
&+&B^{'} \big(Tr M\big)^2+C^{'} Tr M Tr M^3+D^{'}\big(Tr M\big)^4+A^{'}Tr M^2 \big(Tr M\big)^2.
\end{eqnarray}
As we have shown in this article, the coefficients $v_{2,1}$, $v_{4,1}$ and $v_{2,2}$ are given by the following two competing calculations found in equation (\ref{theirs}) (Model I) and  (\ref{our}) (Model II):
\begin{eqnarray}
&&~v_{2,1}=-1~,~v_{4,1}=\frac{3}{2}~,~v_{2,2}=0~,~{\rm Model}~{\rm I}\nonumber\\
&&~v_{2,1}=+1~,~v_{4,1}=0~,~v_{2,2}=\frac{1}{8}~,~{\rm Model}~{\rm II}.\label{comp}
\end{eqnarray}
Explicitly we have
\begin{eqnarray}
&&{\rm Model}~{\rm I}:\nonumber\\
&&V=Tr\big(B M^2+C M^4\big)+D\big(Tr M^2\big)^2+B^{'} \big(Tr M\big)^2+C^{'} Tr M Tr M^3\nonumber\\
&&B=b-\frac{aN^2}{2}~,~C=c+\frac{a^2N^3}{4}~,~B^{'}=\frac{aN}{2}~,~C^{'}=-a^2N^2~,~D=\frac{3a^2N^2}{4}.
\end{eqnarray}
\begin{eqnarray}
&&{\rm Model}~{\rm II}:\nonumber\\
&&V=Tr\big(B M^2+C M^4\big)+D\big(Tr M^2\big)^2+B^{'} \big(Tr M\big)^2+D^{'}\big(Tr M\big)^4+A^{'}Tr M^2 \big(Tr M\big)^2\nonumber\\
&&B=b+\frac{aN^2}{2}~,~C=c~,~B^{'}=-\frac{aN}{2}~,~D^{'}=-\frac{a^2}{12}~,~A^{'}=\frac{a^2N}{6}~,~D=-\frac{a^2N^2}{12}.
\end{eqnarray}
There are two independent parameters in these models which we take to be the usual ones $B$ and $C$. It is found that the scaling of the parameters in Monte Carlo is given approximately by
\begin{eqnarray}
\tilde{B}=BN^{-3/2}~,~\tilde{C}=CN^{-2}~,~\tilde{D}=DN^{-1}~,~{\rm etc}.
\end{eqnarray} 
Since only two of these parameters are independent we must choose $\tilde{a}$, for the consistency of the large $N$ limit, to be any fixed number. We then choose for simplicity $\tilde{a}=1$ or equivalently $D=-2\eta N/3$.

These models can be simulated using the ordinary Metropolis algorithm applied to the eigenvalues of  the matrix $M$, i.e. we diagonalize the matrix $M$, add to the above action the contribution coming from the Vandermonde determinant and then simulate the resulting effective action. This method is free from ergodic problems. 

Our first test for the validity of this algorithm, or any other algorithm for that matter, is to look at the Schwinger-Dyson identity given for the above multitrace matrix models  by
\begin{eqnarray}
<\big(2b Tr M^2+4c Tr M^4+2V_2+4V_4\big)>=N^2.
\end{eqnarray}
The second powerful test is to look at the conventional quartic matrix model with $a=0$, viz $V=V_0$. The eigenvalues distributions in the two stable phases (disorder(one-cut) and non-uniform order (two-cut)) as well as the demarcation of their boundary and the behavior of the specific heat across the transition are all well known analytically given respectively by the formulas (\ref{disorder}), (\ref{nonuniform}), (\ref{cr}) and (\ref{CV1}) and (\ref{CV})  below.

\begin{eqnarray}
\rho(\lambda)&=&\frac{1}{N\pi}(2C\lambda^2+B+Cr^2)\sqrt{r^2-\lambda^2}~,~r^2=\frac{1}{3C}(-B+\sqrt{B^2+12 NC}).\label{disorder}
\end{eqnarray}

\begin{eqnarray}
\rho(\lambda)&=&\frac{2C|\lambda|}{N\pi}\sqrt{(\lambda^2-r_{-}^2)(r_{+}^2-\lambda^2)}~,~r_{\mp}^2=\frac{1}{2C}(-B\mp 2\sqrt{NC}).\label{nonuniform}
\end{eqnarray}

\begin{eqnarray}
B_c^2=4NC \leftrightarrow B_c=-2\sqrt{NC}.\label{cr}
\end{eqnarray}

\begin{eqnarray}
&&\frac{C_v}{N^2}=\frac{1}{4}~,~\bar{B}=\frac{B}{B_c}<-1.\label{CV1}
\end{eqnarray}
\begin{eqnarray}
&&\frac{C_v}{N^2}=\frac{1}{4}+\frac{2\bar{B}^4}{27}-\frac{\bar{B}}{27}(2\bar{B}^2-3)\sqrt{\bar{B}^2+3}~,~\bar{B}>-1.\label{CV}
\end{eqnarray}

\subsection{Monte Carlo Tests of Multitrace Approximations}

 The quartic multitrace approximations can be tested and verified directly in Monte Carlo in order to resolve the ambiguity in the coefficients $v$ given in equation (\ref{comp}). We must have as identity the two equations 
\begin{eqnarray}
<a \int dU Tr [L_a,U\Lambda U^{-1}]^2>_{V_0}=<-V_2(\Lambda)>_{V_0}.
\end{eqnarray}
\begin{eqnarray}
<\frac{1}{2}\bigg(a \int dU Tr [L_a,U\Lambda U^{-1}]^2\bigg)^2>_{V_0}=<-V_4(\Lambda)+\frac{1}{2}V_2^2(\Lambda)>_{V_0}.
\end{eqnarray}
The coefficients $v$ appear in the potentials $V_2$ and $V_4$. The expectation values are computed with respect to the conventional quartic matrix model $V_0=V_0(\Lambda)$. 

This test clearly requires the computation of the kinetic term and its square which means in particular that we need to numerically perform the integral over $U$ in the term $\int dU Tr [L_a,U\Lambda U^{-1}]^2$ which is not obvious how to do in any direct way. Equivalently, we can undo the diagonalization in the terms involving the kinetic term to obtain instead the equations
\begin{eqnarray}
<a Tr [L_a,M]^2>_{V_0}=<-V_2>_{V_0}.\label{id1}
\end{eqnarray}
\begin{eqnarray}
<\frac{1}{2}\bigg(a Tr [L_a,M]^2\bigg)^2>_{V_0}=<-V_4+\frac{1}{2}V_2^2>_{V_0}.\label{id2}
\end{eqnarray}
Now the expectation values in the left hand side must be computed with respect to the conventional quartic matrix model $V_0=V_0(M)$ with the full matrix $M=U\Lambda U^{-1}$ instead of the eigenvalues matrix $\Lambda$, i.e. the eigenvalues+angles. The expectation values in the right hand side can be computed either ways.

In other words, the eigenvalues Metropolis algorithm discussed above, which can compute terms such as $<-V_2>_{V_0}$ and $<-V_4+V_2^2/2>_{V_0}$, can not be used to compute the terms $<a Tr [L_a,M]^2>_{V_0}$ and $<\big(a Tr [L_a,M]^2\big)^2/2>_{V_0}$. We use instead the hybrid Monte Carlo algorithm to compute these terms as well as the terms  $<-V_2>_{V_0}$ and $<-V_4+V_2^2/2>_{V_0}$ in order to verify the above equations. This also should be viewed as a counter check for the hybrid Monte Carlo algorithm since we can compare the values of $<-V_2>_{V_0}$ and $<-V_4+V_2^2/2>_{V_0}$ obtained using the hybrid Monte Carlo with those obtained using our eigenvalues Metropolis algorithm. We note, in passing, that the Metropolis algorithm employed for the eigenvalues problem here is far more efficient than the hybrid Monte Carlo applied to the same problem without diagonalization. 

\begin{figure}[htbp]
\begin{center}
\includegraphics[width=5.0cm,angle=-90]{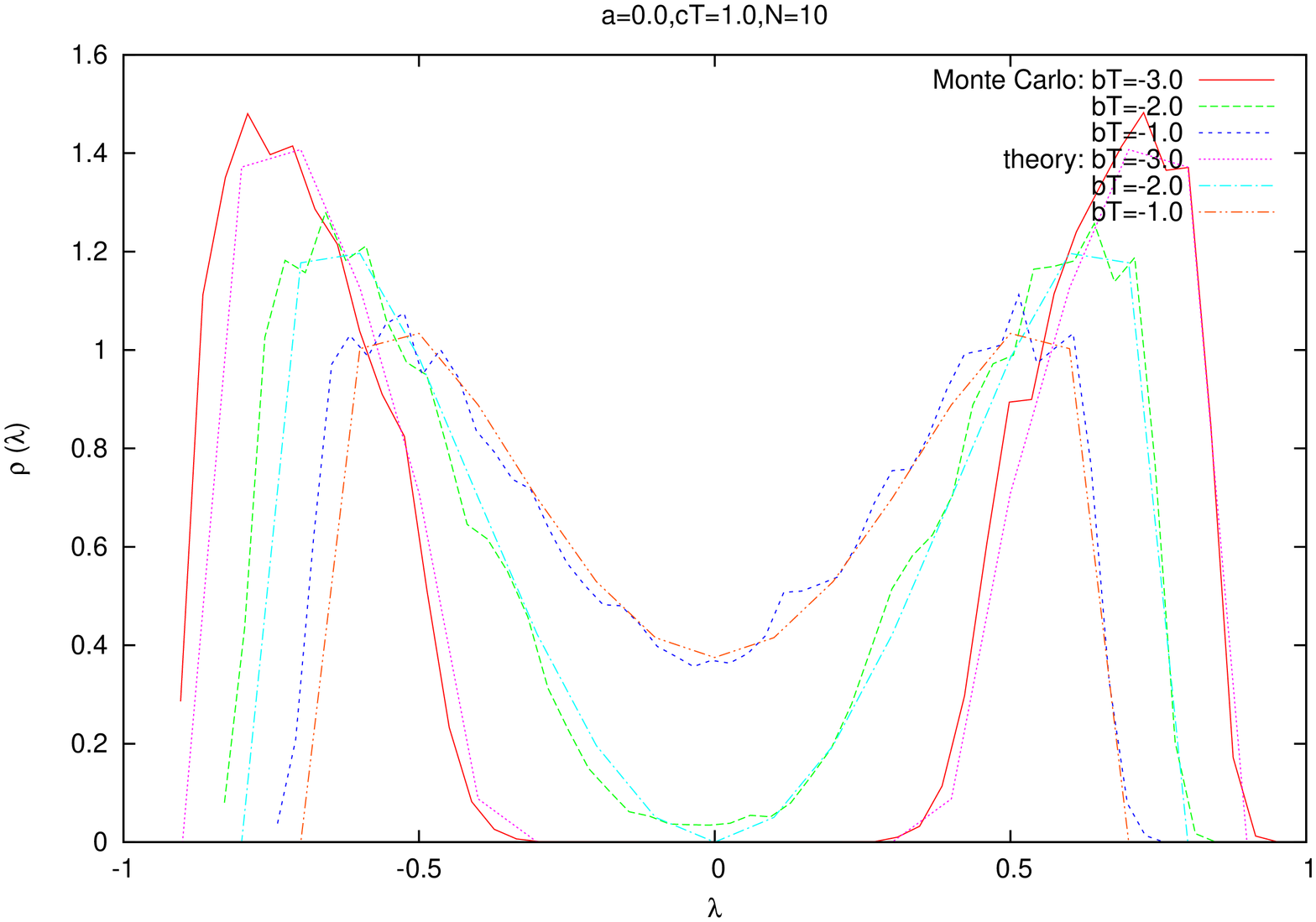}
\includegraphics[width=5.0cm,angle=-90]{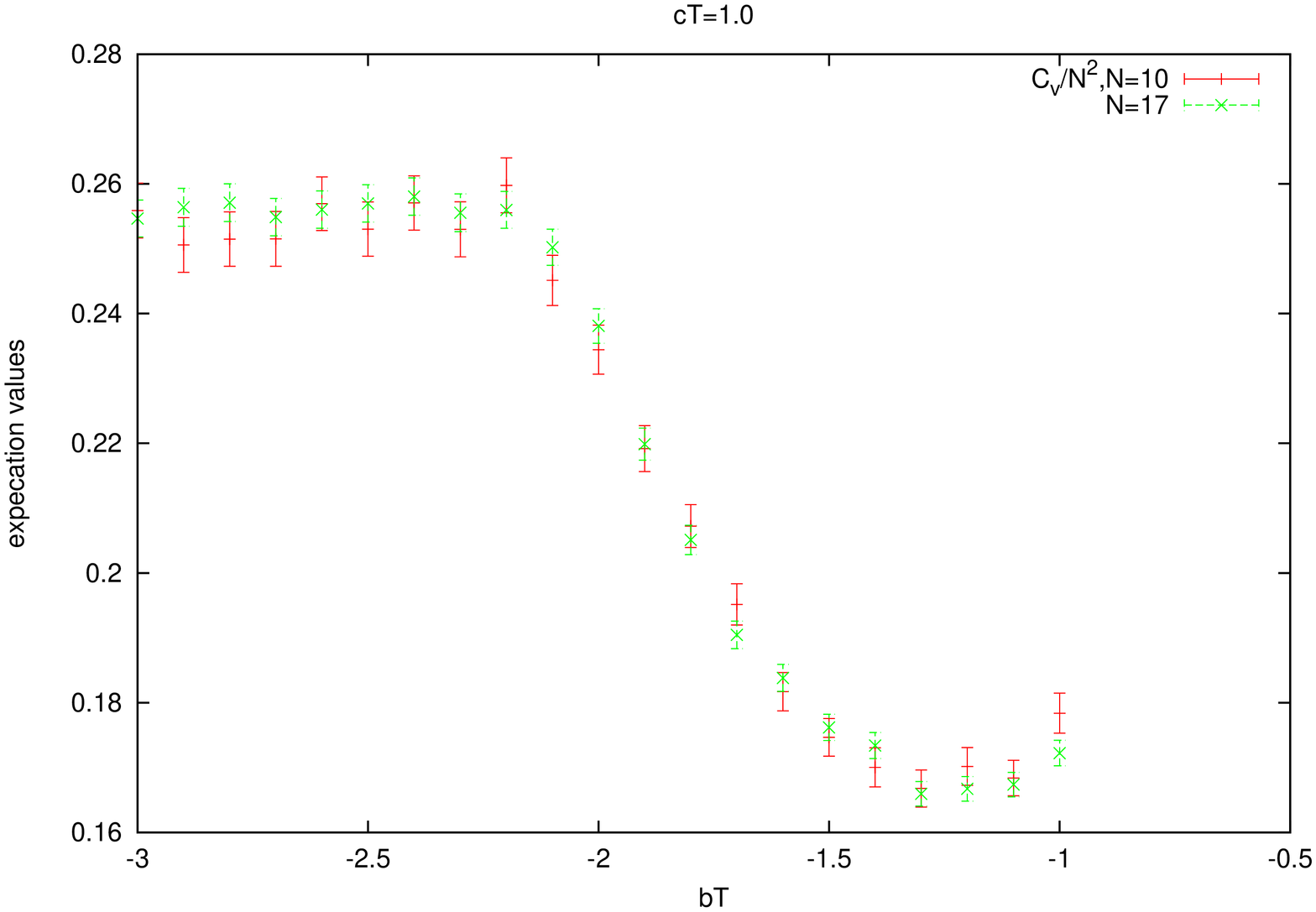}
\caption{The eigenvalues distributions and the specific heat of the quartic matrix model. These are used to calibrate the algorithm against known exact solutions. 
}\label{calibartion}
\end{center}
\end{figure}

In summary, we need to show that the two equations (\ref{id1}) and (\ref{id2}) hold as identities in the correct calculation. In order to solve this problem we need to Monte Carlo sample, both the eigenvalues and the angles of the matrix $M$,  using the hybrid Monte Carlo the quartic matrix model
\begin{eqnarray}
V_0&=&{b}Tr M^2+{c}Tr M^4.
\end{eqnarray}
Clearly, we can choose without any loss of generality $c$ such that $\tilde{c}=1$. Monte Carlo simulations of this model can also be compared to the exact solution outlined in the previous subsection so calibration in this case is easy. A sample of this calculation including the eigenvalues distributions and the specific heat across the transition point are shown on figure (\ref{calibartion}). We can be satisfied from these results that the algorithm and simulations are working properly.

\begin{figure}[htbp]
\begin{center}
\includegraphics[width=5.0cm,angle=-90]{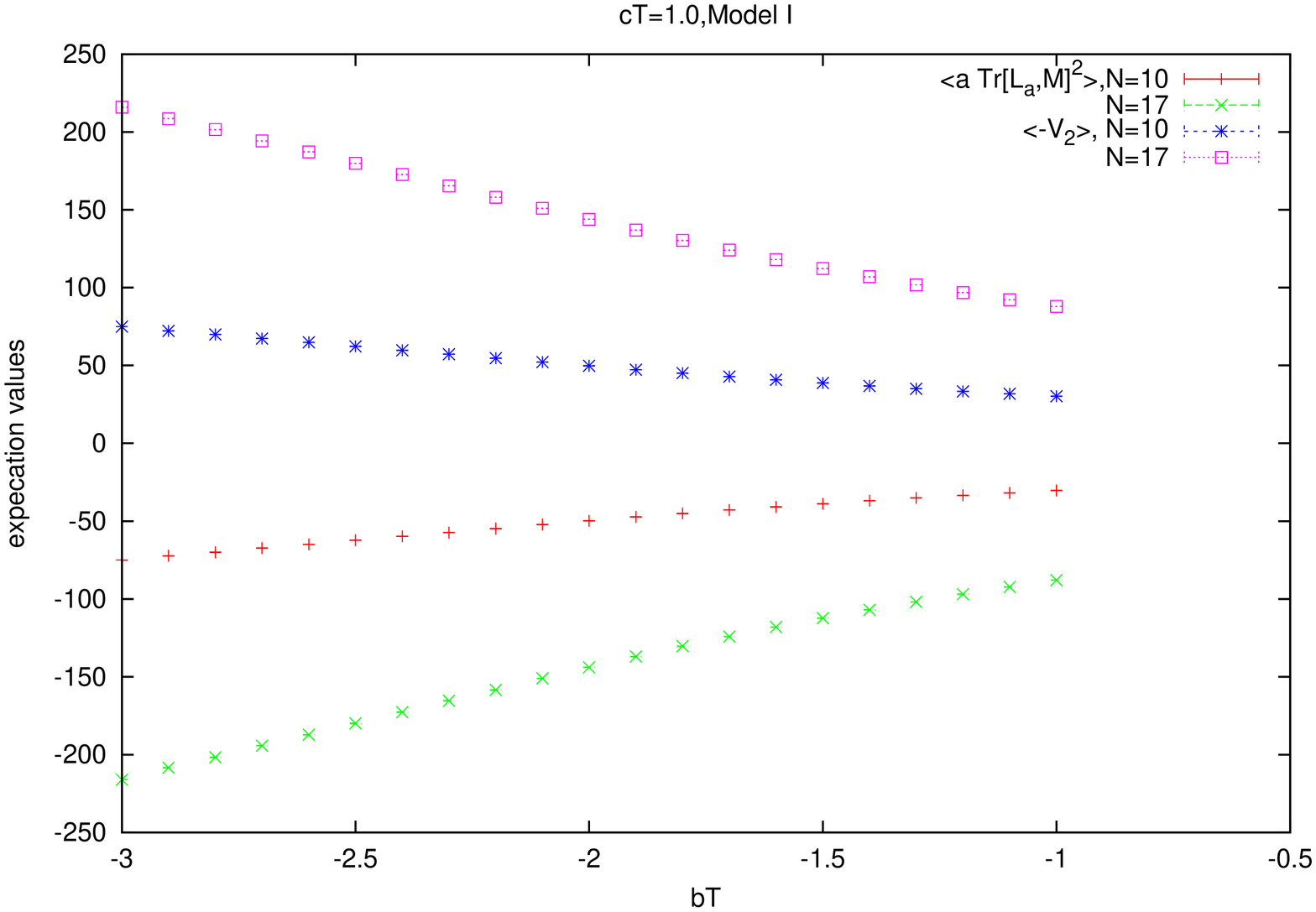}
\includegraphics[width=5.0cm,angle=-90]{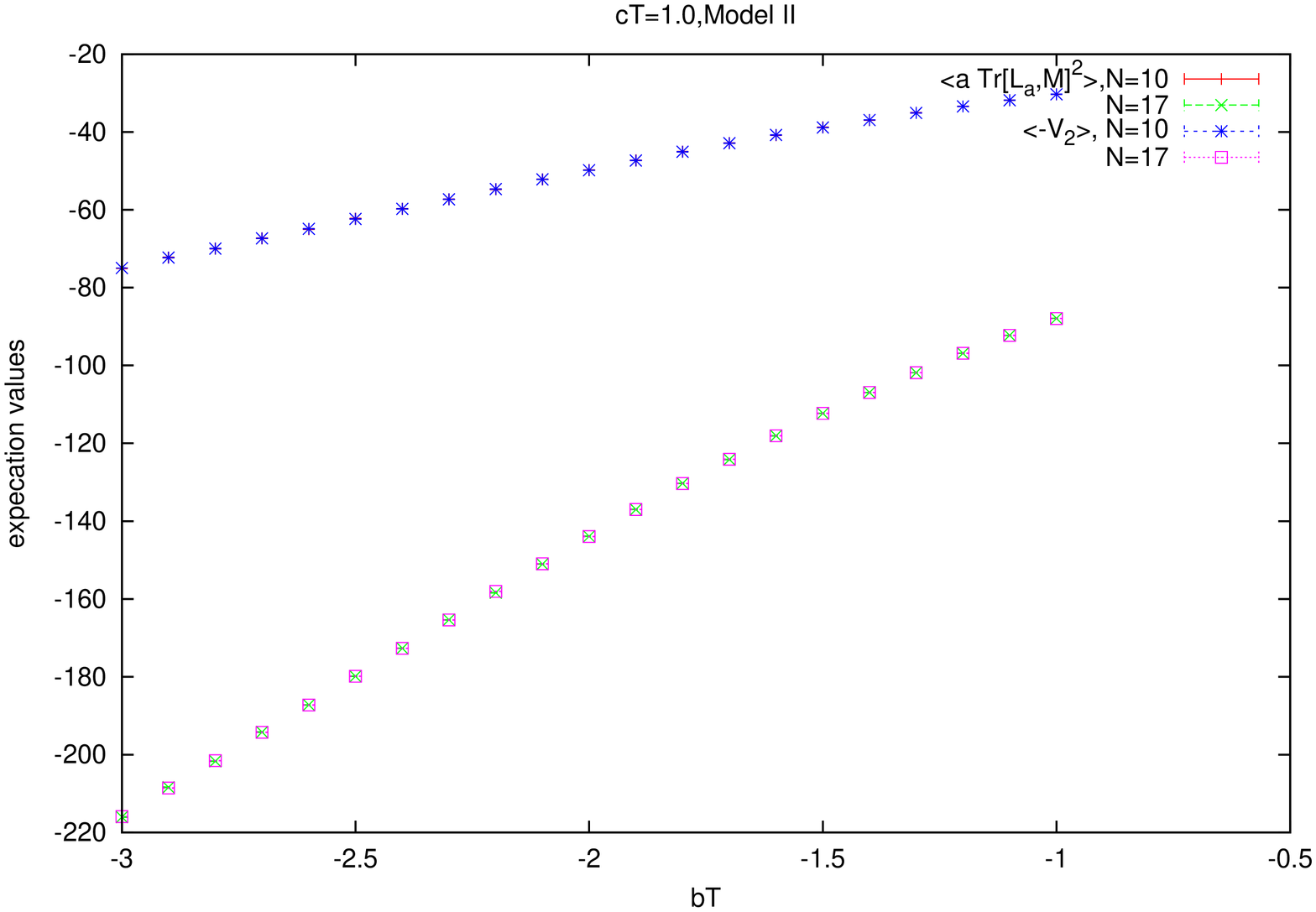}
\includegraphics[width=5.0cm,angle=-90]{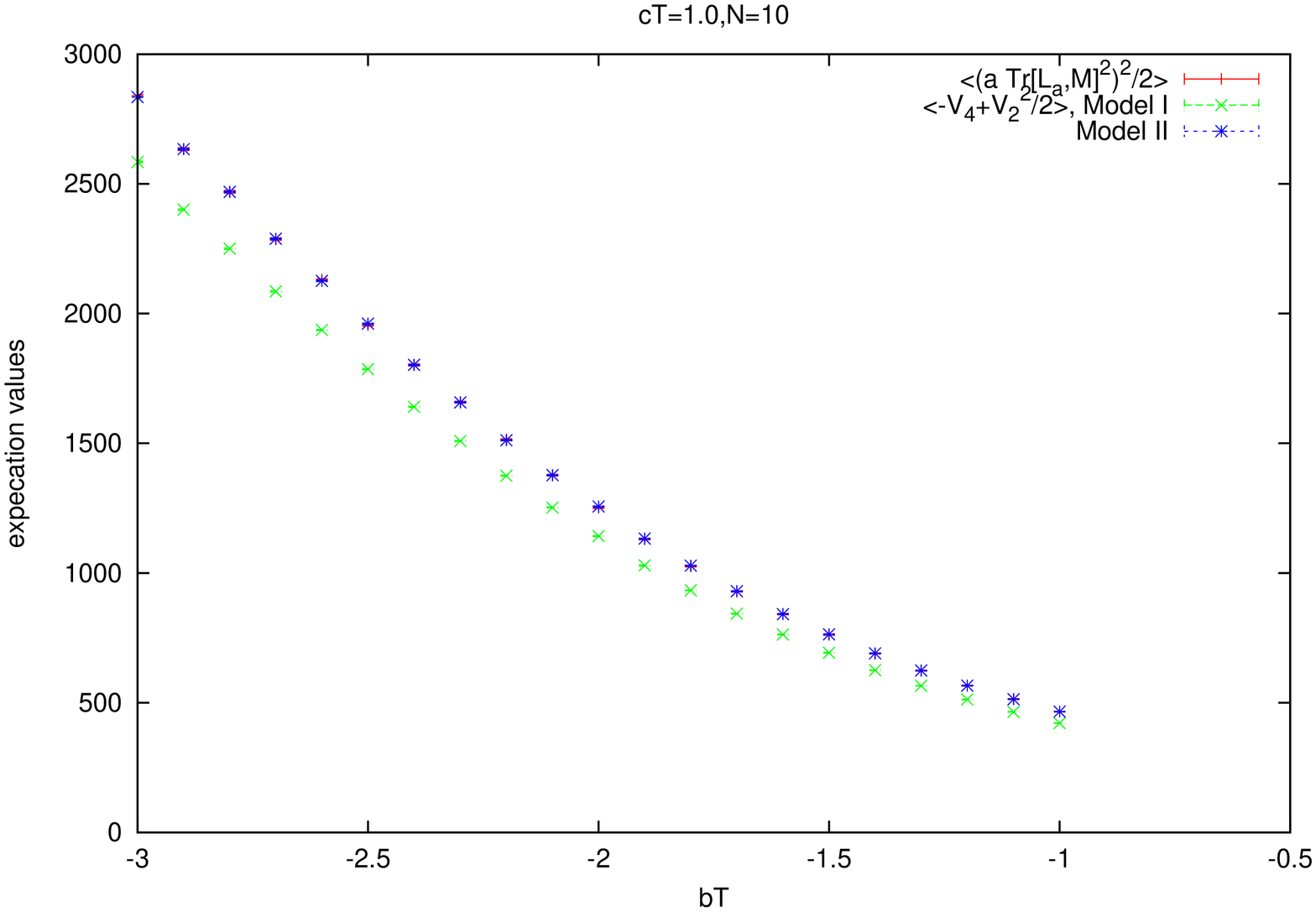}
\includegraphics[width=5.0cm,angle=-90]{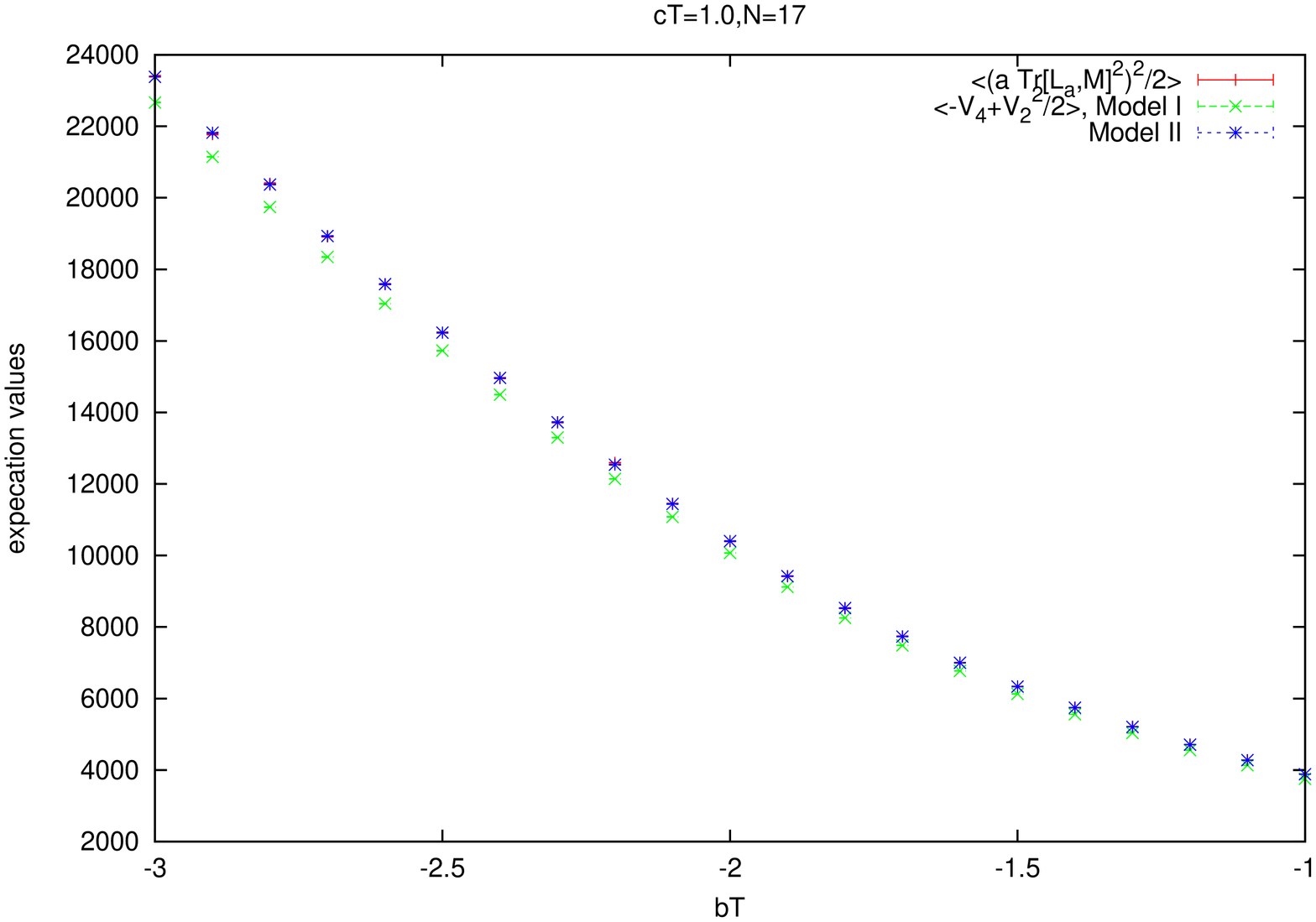}
\caption{The behaviors of multitrace matrix models I and II against the behavior of noncommutative scalar phi-four on the fuzzy sphere. 
}\label{test}
\end{center}
\end{figure}

The two identities (\ref{id1}) and (\ref{id2}) are shown on figure (\ref{test}) for $N=10$ and $N=17$ with $\tilde{c}=1$. It is decisively shown that the calculation of  the coefficients $v$ reported in this article (Model II) gives the correct approximation of noncommutative scalar $\Phi_2^4$ on the fuzzy sphere. Indeed, The data points for the expectation values  $<-V_2>_{V_0}$ and $<-V_4+V_2^2/2>_{V_0}$ in model II coincide, within the best statistical errors, with the data points of the kinetic terms  $<a Tr [L_a,M]^2>_{V_0}$ and $<\big(a Tr [L_a,M]^2\big)^2/2>_{V_0}$ respectively. The discrepancy with model II is obvious on figure  (\ref{test}).

\subsection{Phase Diagrams and Other Physics}

We can now turn to the more serious study of the phase diagrams, critical boundaries, triple point and critical exponents of the multitrace matrix models I and II using Monte Carlo. This is a long calculation which can only be reported elsewhere \cite{ar1,ar2}.  Here we summarize some of our results which include:
\begin{itemize}
\item The phase diagram of model I contains three stable phases: i) disordered (symmetric, one-cut, disk) phase, ii) uniform ordered (Ising, broken, asymmetric one-cut) phase and iii) non-uniform ordered (matrix, stripe, two-cut, annulus) phase which meet at a triple point. The non-uniform ordered phase is a full blown nonperturbative manifestation of the perturbative  UV-IR mixing effect  which is due to the underlying highly non-local matrix degrees of freedom of the noncommutative scalar field. The  critical boundaries are determined and the triple point is located.
\item The uniform ordered phase exists in the model I only with the odd terms included. If we assume the symmetry $M\longrightarrow -M$ then all odd terms can be set to zero and the uniform ordered phase disappears. This is at least true in the domain studied in this article which includes the triple point of fuzzy $\Phi^4$ on the fuzzy sphere and extends to all its phase diagram probed in \cite{GarciaFlores:2009hf,GarciaFlores:2005xc}.
\item The delicate computation of the critical exponents of the Ising transition is discussed and our estimate of the critical exponents $\nu,\alpha,\gamma,\beta$ agrees very well with the Onsager values.
\item  The phase diagram of model II, with or without odd terms, does not contain the uniform ordered phase.
\item The one-cut-to-two-cut transition line does not extend to the origin  in the model II which gives us an estimation of the triple point in this case. 
\item As we have shown in this article, in model II without odd terms, the termination point can be computed from the requirement that the critical point $\tilde{B}_*$ remains always negative. The result is given in equation  (\ref{term}) which agrees with what obtain in Monte Carlo. 
\item In model II with odd terms the termination point is found numerically to be located at  $(\tilde{B},\tilde{C})=(-1.05,0.4)$. This is our measurement of the triple point.
\item In all cases the one-cut-to-two-cut matrix transition line agrees better with the doubletrace matrix theory, studied in this article, than with the quartic matrix model. We recall that the  doubletrace matrix theory is given by $D\neq0$ while all primed parameters are zero.
\item The model of Grosse-Wulkenhaar can also be discussed along the same lines using a combination of the multitrace approach and Monte Carlo approach.
\end{itemize}
We note in passing that other far more important physics can also be obtained from these multitrace matrix models  \cite{ar2}. More precisely, a novel scenario for the emergence of geometry in generic random multitrace matrix models of a single hermitian matrix $M$ with unitary $U(N)$ invariance, i.e. without kinetic term, can be formulated as follows. If the multitrace matrix model under consideration does not sustain the uniform ordered phase then there is no emergent geometry. On the other hand, if  the uniform ordered phase  is sustained then there is an underlying or emergent geometry with dimension determined  from the critical exponents of the uniform-to-disordered (Ising) phase transition and a metric (Laplacian, propagator) determined from the Wigner semicircle law behavior of the eigenvalues distribution of the matrix $M$. 

\section{The Nonperturbative Effective Potential Approach}
The formalism due to  Nair, Polychronakos and Tekel \cite{Polychronakos:2013nca,Tekel:2014bta,Nair:2011ux,Tekel:2013vz} will allow us to compute  the even part of the nonperturbative effective potential, i.e. the part of the potential symmetric under $M\longrightarrow -M$, as a multitrace matrix model.  This will also allow us to compare our multitrace matrix models obtained here, at least in this special case, to an independent exact result. We slightly change notation and start with the action 
\begin{eqnarray}
S=Tr\big(\frac{1}{2}rM^2+gM^4\big).
\end{eqnarray}
We define the moments $m_n$ as usual by $m_n=Tr M^n=\sum_ix_i^n$. By assuming that the kinetic operator ${\cal K}$ satisfies ${\cal K}({\bf 1})=0$ and that odd moments are zero we get immediately 
\begin{eqnarray}
\int dU \exp\big(-\frac{1}{2}Tr M{\cal K}M\big)=\exp(-S_{\rm eff}(t_{2n}))~,~t_{2n}=Tr\big(M-\frac{1}{N}TrM\big)^{2n}.
\end{eqnarray}
Let us first consider the free theory $g=0$. In the limit $N\longrightarrow\infty$ we know that planar diagrams dominates and thus the eigenvalues distribution of $M$, obtained via the calculation of $Tr M^n$, is a Wigner semicircle law  \cite{Steinacker:2005wj,Nair:2011ux}
\begin{eqnarray}
\rho(x)=\frac{2N}{\pi R_W^2}\sqrt{R_W^2-x^2}~,~
R_W^2=\frac{4f(r)}{N}~,~f(r)=\sum_{l=0}^{N-1}\frac{2l+1}{{\cal K}(l)+r}.\label{rw1}
\end{eqnarray}
We consider now $g\neq 0$. The equation of motion of the eigenvalue $x_i$ arising from the effective action $S_{\rm eff}$ contains a linear term in $x_i$ $+$ the Vandermonde contribution $+$ higher order terms. Explicitly, we have
\begin{eqnarray}
\sum_n\frac{\partial S_{\rm eff}}{\partial t_{2n}}2n x_i^{2n-1}=2\sum_{i\neq j}\frac{1}{x_i-x_j}.
\end{eqnarray}
The semicircle distribution is a solution for $g\neq 0$ since it is a solution for $g=0$ \cite{Polychronakos:2013nca}. The term $n=1$ alone will give the semicircle law. Thus the terms $n>1$ are cubic and higher order terms which cause the deformation of the semicircle law. These terms must vanish when evaluated on the semicircle distribution in order to guarantee that the semicircle distribution remains a solution. We rewrite the action  $S_{\rm eff}$ as the following power series in the eigenvalues  

 \begin{eqnarray}
S_{\rm eff}&=&a_2t_2+(a_4t_4+a_{22}t_2^2)+(a_6t_6+a_{42}t_4t_2+a_{222}t_2^3)\nonumber\\
&+&(a_8+a_{62}t_6t_2+a_{422}a_4t_2^2+a_{2222}t_2^4)+...
\end{eqnarray}
We impose then the condition 
 \begin{eqnarray}
\frac{\partial S_{\rm eff}}{\partial t_{2n}}|_{{\rm Wigner}}&=&0~,~n>1,
\end{eqnarray}
and use the fact that the moments in the Wigner distribution satisfy
 \begin{eqnarray}
t_{2n}=C_nt^n~,~C_n=\frac{(2n)!}{n!(n+1)!},
\end{eqnarray}
to get immediately the conditions
\begin{eqnarray}
a_4=0~,~a_6=a_{42}=0~,~a_8=a_{62}=0~,~4a_{44}+a_{422}=0~,~....
\end{eqnarray}
By plugging these values back into the effective action we obtain the form
 \begin{eqnarray}
S_{\rm eff}&=&\frac{1}{2}F(t_2)+(b_1+b_2t_2)(t_4-2t_{2}^{2})^2+c(t_6-5t_2^3)(t_4-3t_2^2)+...
\end{eqnarray}
Thus the effective action is still an arbitrary function $F(t_2)$ of $t_2$ but it is fully fixed in the higher moments $t_4$, $t_6$,.... We note that the extra terms vanish for the Wigner semicircle law. The action up to $6$ order in the eigenvalues depends therefore only on $t_2$, viz  
  \begin{eqnarray}
S_{\rm eff}&=&\frac{1}{2}F(t_2)+...
\end{eqnarray}
The equations of motion of the eigenvalues for $g=0$ read now explicitly 
\begin{eqnarray}
(F^{'}(t_2)+r)x_i=2\sum_{i\neq j}\frac{1}{x_i-x_j}.
\end{eqnarray}
The radius of the semicircle distribution is immediately obtained by
\begin{eqnarray}
R_W^2=\frac{4N}{F^{'}(t_2)+r}.\label{rw2}
\end{eqnarray}
By comparing (\ref{rw1}) and (\ref{rw2}) we obtain the self-consistency equation
 \begin{eqnarray}
\frac{4f(r)}{N}=\frac{4N}{F^{'}(t_2)+r}.
\end{eqnarray}
Another self-consistency condition is the fact that $t_2$ computed using the effective action $ S_{\rm eff}$ for $g=0$, i.e. using the Wigner distribution,  should give the same value, viz
\begin{eqnarray}
t_2&=&Tr M^2=\int_{-R_W}^{R_W} dx x^2\rho(x)=\frac{N}{4}R_W^2=\frac{N^2}{F^{'}(t_2)+r}.
\end{eqnarray}
We have then the two conditions 
\begin{eqnarray}
F^{'}(t_2)+r=\frac{N^2}{t_2}~,~t_2=f(r).
\end{eqnarray}
The solution is given by
\begin{eqnarray}
F^{}(t_2)=N^2\int dt_2 (\frac{1}{t_2}-\frac{1}{N^2} g(t_2)).
\end{eqnarray}
$g(t_2)$ is the inverse function of $f(r)$, viz $f(g(t_2))=t_2$.

For the case of the fuzzy sphere with a kinetic term given by the canonical  formula ${\cal K}(l)=l(l+1)$ we have the result
\begin{eqnarray}
f(r)=\ln \big(1+\frac{N^2}{r}\big).
\end{eqnarray}
Thus the corresponding solution is explicitly given by
\begin{eqnarray}
F(t_2)=N^2\ln\frac{t_2}{1-\exp(-t_2)}.
\end{eqnarray}
The full effective action on the sphere is then
 \begin{eqnarray}
S_{\rm eff}&=&\frac{N^2}{2}\ln\frac{t_2}{1-\exp(-t_2)}+Tr\big(\frac{1}{2}rM^2+gM^4\big)+...\nonumber\\
&=&\frac{N^2}{2}\bigg(\frac{t_2}{2}-\ln\frac{\exp(t/2)-\exp(-t/2)}{t}\bigg)+Tr\big(\frac{1}{2}rM^2+gM^4\big)+...\nonumber\\
&=&\frac{N^2}{2}\bigg(\frac{t_2}{2}-\frac{1}{24}t_2^2+\frac{1}{2880}t_2^4+...\bigg)+Tr\big(\frac{1}{2}rM^2+gM^4\big)+...\label{poly}
\end{eqnarray}
This should be compared with our result in this article with action given by $aTr M{\cal K}M+b Tr M^2+c Tr M^4$ and effective action given by our equation (\ref{our})  or equivalently 
 \begin{eqnarray}
V_0+\Delta V_0
&=&\bigg(\frac{aN^2}{2}Tr M^2 -\frac{a^2N^2}{12}(Tr M^2)^2+...\bigg)+Tr\big(bM^2+cM^4\big)+...\nonumber\\
\end{eqnarray}
It is very strange that  the author of \cite{Polychronakos:2013nca} notes that their result (\ref{poly}) is in agreement with the result of \cite{O'Connor:2007ea}, given by equation $(4.5)$, which involves the term $T_4=\sum_{i\neq j}(x_i-x_j)^4/2$. It is very clear that $T_4$ is not present in the above equation (\ref{poly}) which depends  instead on the term $T_2^2$ where  $T_2=\sum_{i\neq j}(x_i-x_j)^2/2$. The work \cite{Saemann:2010bw} contains the correct calculation which agrees with both the results of  \cite{Polychronakos:2013nca} and our result here.

The one-cut-to-two-cut phase transition derived from the effective action $S_{\rm eff}$ will be appropriately shifted. The equation determining the critical point is still given, as before, by the condition that the eigenvalues distribution becomes negative. We get \cite{Polychronakos:2013nca}
\begin{eqnarray}
r=-5\sqrt{g}-\frac{1}{1-\exp(1/\sqrt{g})}.
\end{eqnarray}
For large $g$ we obtain
\begin{eqnarray}
r=-\frac{1}{2}-4\sqrt{g}+\frac{1}{12\sqrt{g}}+....
\end{eqnarray}
This is precisely the result obtained in  this article given by equation (\ref{our2}) with the identification $a=1$, $b=r$ and $c=4g$.


\section{Conclusion}
In this article we have extended the multitrace approach of \cite{O'Connor:2007ea} to two-dimensional noncommutative phi-four theory with non-zero harmonic oscillator term on the fuzzy sphere and on the Moyal-Weyl plane. We computed the corresponding real multitrace quartic matrix model  upto the second order in the kinetic term parameter then derived explicitly, in the case of the even doubletrace matrix models, the critical transition line between the one-cut (disordered,disk) phase with $<\Phi>=0$ and the two-cut (non-uniform ordered,annulus) phase with $<\Phi>=\gamma$. A robust prediction  of the triple point, identified as a termination point of the matrix transition line, is  derived and compared with the previous Monte Carlo result of \cite{GarciaFlores:2009hf,GarciaFlores:2005xc}.  Our estimation is improved considerably by including odd moments in the effective multitrace action as evidenced by Monte Carlo simulations of this multitrace matrix model.

The multitrace matrix model of  \cite{O'Connor:2007ea} as well as the one obtained in this article are tested for their correctness using Monte Carlo where it is decisively shown that our calculation here gives the correct approximation of noncommutative scalar $\Phi_2^4$ on the fuzzy sphere to the second order. This was also confirmed using the nonperturbative multitrace approach of \cite{Polychronakos:2013nca,Tekel:2014bta,Nair:2011ux,Tekel:2013vz}. The rich phase diagrams of both of these models, obtained in Monte Carlo simulations, are also described in some detail together with some unexpected physics, such as emergent geometry, of generic multitrace matrix models. 



\paragraph{Acknowledgment:}  
This research was supported by CNEPRU: "The National (Algerian) Commission for the Evaluation of
University Research Projects"  under contract number ${\rm DO} 11 20 13 00 09$.

\appendix
\section{Large $N$ Behavior}
The coefficients $s$ are defined, in terms of the kinetic matrix $K$ and characters and dimensions of various SU(N)/U(N) representations, by the following equations
\begin{eqnarray}
s_{1,2}=\frac{1}{{\rm dim}(1,2)}K_{AB}Tr_{(1,2)}t_A\otimes t_B~,~s_{2,1}=\frac{1}{{\rm dim}(2,1)}K_{AB}Tr_{(2,1)}t_A\otimes t_B.
\end{eqnarray}
\begin{eqnarray}
s_{1,4}=\frac{1}{{\rm dim}(1,4)}K_{AB}K_{CD}Tr_{(1,4)}t_A\otimes t_B\otimes t_C\otimes t_D.
\end{eqnarray}
\begin{eqnarray}
s_{4,1}=\frac{1}{{\rm dim}(4,1)}K_{AB}K_{CD}Tr_{(4,1)}t_A\otimes t_B\otimes t_C\otimes t_D.
\end{eqnarray}
\begin{eqnarray}
s_{2,3}=\frac{1}{{\rm dim}(2,3)}(2K_{AB}K_{CD}+K_{AD}K_{BC})Tr_{(2,3)}t_A\otimes t_B\otimes t_C\otimes t_D.
\end{eqnarray}
\begin{eqnarray}
s_{3,2}=\frac{1}{{\rm dim}(3,2)}(K_{AB}K_{CD}+2K_{AC}K_{BD})Tr_{(3,2)}t_A\otimes t_B\otimes t_C\otimes t_D.
\end{eqnarray}
\begin{eqnarray}
s_{2,2}=\frac{1}{{\rm dim}(2,2)}(K_{AB}K_{CD}+K_{AC}K_{BD})Tr_{(2,2)}t_A\otimes t_B\otimes t_C\otimes t_D.
\end{eqnarray}
On the other hand, the combinations which appear in the quartic part of the effective potential are given by the following expressions  
\begin{eqnarray}
\frac{1}{8N}(s_{1,4}-s_{4,1}-s_{2,3}+s_{3,2})&=&\frac{1}{16N^6}(-2-\frac{25}{N^2}+O_4)K_{ii,jj}^2+\frac{1}{8N^5}(1+\frac{15}{N^2}+O_4)K_{ii,kl}K_{jj,lk}.\nonumber\\
\end{eqnarray}
\begin{eqnarray}
\frac{1}{6}(s_{1,4}+s_{4,1}-s_{2,2})&=&\frac{1}{4N^6}(2+\frac{25}{N^2}+O_4)K_{ii,jj}^2-\frac{1}{2N^5}(1+\frac{15}{N^2}+O_4)K_{ii,kl}K_{jj,lk}\nonumber\\
&-&\frac{1}{4N^5}(1+\frac{15}{N^2}+O_4)K_{ii,jj}K_{kl,lk}+\frac{1}{4N^4}(1+\frac{17}{N^2}+O_4)K_{ij,jl}K_{kk,li}.\nonumber\\
\end{eqnarray}
\begin{eqnarray}
\frac{1}{8N}(s_{1,4}-s_{4,1}+s_{2,3}-s_{3,2}-2(s_{1,2}^2-s_{2,1}^2))&=&\frac{1}{8N^6}(-1+O_2)K_{ij,kl}K_{ji,lk}+\frac{1}{16N^6}(-\frac{18}{N^2}+O_4)K_{ii,jj}^2\nonumber\\
&+&\frac{5}{4N^7}(1+O_2)K_{ii,kl}K_{jj,lk}+\frac{1}{16N^5}(\frac{12}{N^2}+O_4)K_{ii,jj}K_{kl,lk}\nonumber\\
&+&\frac{1}{8N^5}(1+O_2)K_{ij,ki}K_{lk,jl}-\frac{3}{4N^6}(1+O_2)K_{ij,jl}K_{kk,li}\nonumber\\
&+&\frac{1}{8N^6}(-1+O_2)K_{ij,ji}^2.
\end{eqnarray}
\begin{eqnarray}
\frac{1}{16N^2}(s_{1,4}+s_{4,1}-s_{2,3}-s_{3,2}+2s_{2,2}-2(s_{1,2}-s_{2,1})^2)&=&\frac{1}{16N^6}(1+O_2)K_{ij,kl}K_{ji,lk}+\frac{1}{32N^6}(\frac{6}{N^2}+O_4)K_{ii,jj}^2\nonumber\\
&-&\frac{1}{8N^7}(2+O_2)K_{ii,kl}K_{jj,lk}.
\end{eqnarray}
\begin{eqnarray}
\frac{1}{48}(s_{1,4}+s_{4,1}+3s_{2,3}+3s_{3,2}+2s_{2,2}-6(s_{1,2}+s_{2,1})^2)
&=&\frac{1}{16N^6}(1+O_2)K_{ij,kl}K_{ji,lk}+\frac{1}{32N^6}(\frac{18}{N^2}+O_4)K_{ii,jj}^2\nonumber\\
&-&\frac{5}{8N^7}(1+O_2)K_{ii,kl}K_{jj,lk}\nonumber\\
&+&\frac{1}{16N^5}(-\frac{8}{N^2}+O_4)K_{ii,jj}K_{kl,lk}\nonumber\\
&-&\frac{1}{8N^5}(1+O_2)K_{ij,ki}K_{lk,jl}+\frac{1}{4N^6}(2+O_2)K_{ij,jl}K_{kk,li}\nonumber\\
&+&\frac{1}{32N^4}(\frac{4}{N^2}+O_4)K_{ij,ji}^2.
\end{eqnarray}
The kinetic matrix $K_{ij,kl}$ is defined in terms of the kinetic matrix $K_{AB}$, which is defined by equation (\ref{KAB}), by 
 \begin{eqnarray}
K_{AB}=(t_A)_{jk}(t_B)_{li}K_{ij,kl}.
\end{eqnarray}
The Large $N$ behavior of the different operators is given by

\begin{eqnarray}
\frac{1}{N^5}K_{ii,kl}K_{jj,lk}&=&\frac{r^4}{3}\bigg(13-10\epsilon \bigg)+...
\end{eqnarray} 
\begin{eqnarray}
K_{ij,kl}K_{li,jk}
&=&16r^4(\sqrt{\omega}+1)^2\bigg(\frac{1}{3}N^3-\epsilon\frac{3}{10}N^3\bigg)+....
\end{eqnarray} 
\begin{eqnarray}
K_{ij,jl}K_{kk,li}
&=&8r^4(\sqrt{\omega}+1)\bigg(\frac{7}{12}N^4-\epsilon\frac{1}{2}N^4\bigg)+....
\end{eqnarray} 
\begin{eqnarray}
K_{ij,kl}K_{ji,lk}&=&\frac{2r^4N^4}{9}(21-16\epsilon)+\frac{2r^4\omega N^4}{9}(9-8\epsilon)+...
\end{eqnarray}
\begin{eqnarray}
\frac{1}{4N^6}K_{ii,jj}^2=r^4\bigg(1-3\frac{\epsilon}{4}\bigg)+....\nonumber\\
\end{eqnarray} 
\begin{eqnarray}
K_{ii,jj}K_{kl,lk}
&=&4r^4(\sqrt{\omega}+1)\bigg(N^5-\epsilon \frac{5}{6}N^5\bigg)+....\nonumber\\
\end{eqnarray} 
\begin{eqnarray}
K_{mj,km}K_{nk,jn}=16 r^4(\sqrt{\omega}+1)^2\bigg(\frac{1}{3}N^3-\epsilon\frac{3}{10}N^3\bigg)+....
\end{eqnarray} 
\begin{eqnarray}
K_{ij,ji}^2
&=&16 r^4(\sqrt{\omega}+1)^2\bigg(\frac{1}{4}N^4-\epsilon\frac{2}{9}N^4\bigg)+....
\end{eqnarray}

\section{Result of \cite{O'Connor:2007ea} Revisited} The starting point is the result given by equation $(3.25)$ of \cite{O'Connor:2007ea} which in our notation reads  (with $a=2\pi/(N+1)$)
\begin{eqnarray}
\int dU~\exp\bigg(aTr[U^{-1}L_a U,{\Lambda}]^2 \bigg)
&=&1+2a.\frac{T_2}{N(N^2-1)}K_{aa}\nonumber\\ 
&+&8a^2.\frac{T_2^2-2T_4}{4N^2(N^2-1)(N^2-9)}X_1\nonumber\\
&+&8a^2.\frac{-5T_2^2+(N^2+1)T_4}{2N(N^2-1)(N^2-4)(N^2-9)}X_2\nonumber\\
&+&....\label{our1}
\end{eqnarray}
In the above equation we have multiplied by the appropriate factor and also included,  for completeness, the first and second order correction terms. We should make the identification $\Xi=T$ between our notation and the notation  of \cite{O'Connor:2007ea}. The operators $X_1$ and $X_2$ are defined by
\begin{eqnarray}
X_1=2K_{ab}^2+K_{aa}^2.
\end{eqnarray} 
\begin{eqnarray}
X_2=K_{ab}K_{cd}(\frac{1}{2}d_{abk}d_{cdk}+d_{adk}d_{bck}).
\end{eqnarray} 
In other words, $X_1$ and $X_2$ are essentially the operators   $2tr K^2+(tr K)^2$ and  $K^{\perp}K$ of \cite{O'Connor:2007ea}.  We must furthermore take into account the different normalizations for the Gell Mann matrices employed in the two cases. The kinetic matrix in this case is defined by
\begin{eqnarray}
K_{ab}=Tr[L_i,t_a][L_i,t_b].
\end{eqnarray}
The source of the discrepancy between our result (\ref{our}) and the result obtained in  \cite{O'Connor:2007ea} was traced to the operator $K^{\perp}K$, i.e. $X_2$, defined in equation $(3.18)$  of \cite{O'Connor:2007ea} which was neglected in the large $N$ limit in their analysis. The operators $X_1$ and $X_2$ can be computed in closed form. We find

\begin{eqnarray}
X_1=\frac{N^4(N^2-1)^2}{16}+\frac{N^2(N^2-1)^2}{6}.
\end{eqnarray}
\begin{eqnarray}
X_2
&=&\frac{N^3(N^2-1)^2}{16}-\frac{N(N^2-1)^2}{6}-\frac{N(N^2-1)}{4}.
\end{eqnarray}
Clearly $X_1$ is of order $N^8$ while $X_2$ is of order $N^7$. As a consequence 
the coefficient of $T_4$  in (\ref{our1}) comes out to be subleading. This can be inferred quite easily from the exact result\footnote{This can be derived directly for $N=2$.}
\begin{eqnarray}
\int dU~\exp\bigg(aTr[U^{-1}L_a U,{\Lambda}]^2 \bigg)
&=&1-\frac{aN}{2}T_2+\frac{a^2}{24}(T_2^2-2T_4)\frac{N^2-1}{N^2-9}(3N^2+8)\nonumber\\
&+&\frac{a^2}{12}(-5T_2^2+(N^2+1)T_4)\frac{3N^2+1}{N^2-9}+....\nonumber\\
\end{eqnarray}
This leads to (\ref{our}). Furthermore,  it is obvious from this result that although the subleading coefficient  of the operator $ T_2^2$ is of order $N^0$, the contribution associated with this term can not be suppressed, since this operator is actually of order $N^4$. 

\section{Quartic Equation}
The quartic equation (\ref{quex}) reads in terms of the scaled parameters $\bar{b}$, $\bar{c}$ and $\bar{x}=\tilde{a}\gamma^{-1/4}x$ as follows

\begin{eqnarray}
&&\bar{x}^4+\alpha \bar{x}^2+\beta\bar{x}+1=0.
\end{eqnarray}
\begin{eqnarray}
&&\alpha=\gamma^{\frac{1}{2}}\bar{\alpha}~,~\beta=\gamma^{\frac{1}{4}}\bar{\beta}.
\end{eqnarray}
\begin{eqnarray}
&&\gamma=\frac{24}{(\Omega^2+1)^2\eta \bar{c}}.
\end{eqnarray}
\begin{eqnarray}
&&\bar{\alpha}=-\frac{1}{12}\big(9\bar{c}-2\eta(\Omega^2+1)^2\big)~,~\bar{\beta}=-\frac{1}{4}\big(2\bar{b}+\Omega^2+1\big).
\end{eqnarray}
The range (\ref{rangec}) of $\bar{c}$ translates to a range of $\bar{\alpha}$ given by
\begin{eqnarray}
\bar{\alpha}\leq -\frac{\eta(\Omega^2+1)^2}{3}.
\end{eqnarray}
The four solutions of our depressed quartic equation can be rewritten as
\begin{eqnarray}
\bar{x}=\frac{1}{2}\bigg[\pm_{1}\sqrt{z}\pm_{2}\sqrt{-\big(z+2\alpha\pm_1\frac{2\beta}{\sqrt{z}}\big)}\bigg],\label{qur}
\end{eqnarray}
where $z$ is a solution of the cubic equation 
\begin{eqnarray}
z^3+2\alpha z^2+(\alpha^2-4)z-\beta^2=0.
\end{eqnarray}
Define
\begin{eqnarray}
z=t-\frac{2\alpha}{3}.\label{qur1}
\end{eqnarray}
The corresponding depressed cubic equation is 
\begin{eqnarray}
t^3+3Q t-2R=0.\label{cubic}
\end{eqnarray}
\begin{eqnarray}
Q=-\frac{\alpha^2}{9}-\frac{4}{3}~,~R=-\frac{4\alpha}{3}+\frac{\alpha^3}{27}+\frac{\beta^2}{2}.\label{qur2}
\end{eqnarray}
Next we reduce to a quadratic equation. We start from the identity 
\begin{eqnarray}
(t^3-B^3)+C(t-B)=(t-B)(t^2+Bt+B^2+C). 
\end{eqnarray}
By comparison we get
\begin{eqnarray}
C=3Q~,~B^3+CB=2R.
\end{eqnarray}
In other words, $B$ solves the cubic equation
\begin{eqnarray}
B^3+3Q B-2R=0.
\end{eqnarray}
The solution is immediately given by
\begin{eqnarray}
 B=[R+\sqrt{Q^3+R^2}]^{1/3}+[R-\sqrt{Q^3+R^2}]^{1/3}. 
\end{eqnarray}
We have then
\begin{eqnarray}
t^3+3Q t-2R=(t-B)(t^2+Bt+B^2+3Q). 
\end{eqnarray}
In other words, $t=B$ is a solution of the cubic equation (\ref{cubic}). The other two solutions solve the quadratic equation $t^2+Bt+B^2+3Q=0$, viz
\begin{eqnarray}
t=\frac{-B\pm \sqrt{-3B^2-12Q}}{2}.
\end{eqnarray}
This can be rewritten also as
\begin{eqnarray}
t=\frac{-B\pm i\sqrt{3}A}{2}.
\end{eqnarray}
\begin{eqnarray}
A=[R+\sqrt{Q^3+R^2}]^{1/3}-[R-\sqrt{Q^3+R^2}]^{1/3}. 
\end{eqnarray}
Define
\begin{eqnarray}
D=Q^3+R^2~,~S=\big[R+\sqrt{D}\big]^{1/3}~,~T=\big[R-\sqrt{D}\big]^{1/3}.\label{cb0}
\end{eqnarray}
In summary the real solutions of interest of our cubic equation are given by 
\begin{eqnarray}
D> 0\Rightarrow t=\Big\{S+T\Big\}.\label{cb1}
\end{eqnarray}
\begin{eqnarray}
D\leq 0\Rightarrow t=\Big\{S+T= {\rm real} (2S)~,~\frac{1}{2}(-1\pm i\sqrt{3})S+\frac{1}{2}(-1\mp i\sqrt{3})T={\rm real}\big((-1\pm i\sqrt{3})S\big)\Big\}.\nonumber\\\label{cb2}
\end{eqnarray}
Our numerical approach is based on the solutions (\ref{qur})-(\ref{qur1})-(\ref{qur2}) and (\ref{cb0})-(\ref{cb1})-(\ref{cb2}).

\end{document}